# Large eddy simulation of a supersonic lifted hydrogen flame with perfectly stirred reactor model

Majie Zhao[1], Zhi X. Chen[2,3], Huangwei Zhang[1,*] and Nedunchezhian Swaminathan[2]

[1]Department of Mechanical Engineering, National University of Singapore, 9 Engineering Drive 1, Singapore 117576, Republic of Singapore
[2]Department of Engineering, University of Cambridge, Cambridge CB2 1PZ, United Kingdom
[3]Robinson College, University of Cambridge, Cambridge CB3 9AN, United Kingdom

## Abstract

Large Eddy Simulation with a Perfectly Stirred Reactor model (LES−PSR) is developed to simulate supersonic combustion with high-enthalpy flow conditions. The PSR model considers the viscous heating and compressibility effects on the thermo-chemical state, through correcting the chemical source term for progress variable and incorporating absolute enthalpy as the control variable for the look-up table. It is firstly validated by using *a priori* analysis of the viscous heating and compressibility effects. Then an auto-igniting hydrogen flame stabilized in supersonic vitiated co-flowing jet is simulated with LES−PSR method. The results show that the shock wave structure, overall flame characteristics, flame−shock interaction and lift-off height are accurately captured. Good agreements of the velocity and mixture fraction statistics with the experimental data are observed. The results also show that the LES−PSR model can predict the mean temperature and mole fractions of major species quite well in both flame induction and stabilization zones. However, there are some under-predictions of temperature RMS by about $100-150$ K, which may be due to the chemical non-equilibrium in the $H_2/O_2$-enriched combustion product of the co-flowing jet. The scatter plots of two probe locations respectively from induction and flame zones show that the respective flame structures in mixture fraction space are captured well. However, the fluctuations of the temperature and species mole fractions are under-predicted in the flame zone. The shock-induced auto-igniting spots are captured by the PSR model. These spots are highly unsteady and play an important role in flame stabilization. It is also shown that the intense reactions are initiated at mixture fractions around the stoichiometry or fuel-lean values, corresponding to local elevated pressure ($1.5-2.0$ atm) due to shock compression. The results also demonstrate that the pressure elevation is shown to have significant effects on the most reactive mixture fraction and shortest ignition delay time.

---

* Corresponding author. E-mail: huangwei.zhang@nus.edu.sg. Tel: +65 6516 2557.

# 1. Introduction

Turbulent supersonic combustion becomes increasingly important nowadays due to the interests in developing high-speed propulsion applications, such as ramjet and scramjet engines [1]. Besides similar flame dynamics in low-speed combustion, such as reactant mixing, ignition and flame stabilization, shock-laden flow fields in supersonic combustion add extra complexities, arising mainly from the effects of shock waves and compressibility, which have further influence on the turbulence-combustion interaction (TCI) [2]. For experimental studies on fundamental physics in supersonic combustion, a huge investment is required and numerous practical difficulties in facility setup are needed to be tackled. Conversely, Large Eddy Simulation (LES) is deemed an alternative research method for supersonic combustion, but accurate and physically sound approaches should be developed, to close the sub-grid scale reaction rate incorporating the shock and compressibility effects [2].

Various sub-grid scale combustion models originally developed for low-Mach number flows have been extended for turbulent supersonic combustion, such as quasi-laminar chemistry (QLC) model [3], Partially Stirred Reactor (PaSR) model [4], Transported Probability Density Function (TPDF) model [5], Linear Eddy Model (LEM) [6], and tabulated chemistry approach or flamelet-type model [7]. The QLC and PaSR models estimate the filtered reaction rate directly with the filtered quantities, thereby neglecting the interactions between combustion and turbulence at the sub-grid scale. The multiple physics of high-speed combustion, such as auto-ignition and shock / flame interactions, can be captured by using the above models only when a fine mesh is used to ensure the sub-grid scale Damkohler number ($Da_{sgs}$) is less than unity [3,4,8]. The TPDF method applies an exact closure for the non-linear reaction rates and therefore shows essential advantages over the QLC and also PaSR model due to its capability to approximate scalar fluctuations with not highly-resolved simulations [9–11]. Moreover, LEM is based on one-dimensional spatial structure of reacting flow, which can be used to capture various physical phenomena in high-speed combustion, such as interactions between combustion, turbulence and shock waves [6,12,13]. Although TPDF and LEM do not require a fine LES mesh (hence $Da_{sgs} \ll 1$), they still incur high computational cost to solve the flame information on the notional fluid particles or in the LEM domain [2].

In the tabulated chemistry approach, the thermo-chemical states are pre-calculated from various prototype flames, e.g., non-premixed laminar flames, one dimensional un-stretched premixed flames or Perfectly Stirred Reactor (PSR). Therefore, the computational cost can be considerably reduced compared to TPDF, LEM and QLC methods. Moreover, the tabulated chemistry approaches based on mixture fraction and progress variable have been widely applied for simulations of low-speed turbulent combustion, such as Flamelet / Progress Variable (FPV) approach [14], Flame Prolongation of ILDM (FPI) approach [15] or Flamelet Generated Manifolds (FGM) [16] and PSR model [17]. In these applications, isobaric and atmospheric conditions are assumed. However, for supersonic reacting flows, the effects of pressure and temperature fluctuations induced by shocks are significant and should be included in the thermo-chemical table.

To this end, some attempts have been made for modelling supersonic combustion with tabulated chemistry method [7,18–21]. For instance, the FPV approach [14] developed for subsonic combustion was extended to compressible reacting flows by Pecnik et al. [18] by using a scaled source term of progress variable equation. LES of hydrogen/air scramjet combustion with the compressible FPV method was used by Cao et al. [21], and good agreement with experimental data is achieved except some over-predictions of heat flux and temperature. Note that only the effect of pressure on the source term of the progress variable is considered, which may be responsible for the over-predictions without considering the temperature effect in Ref. [21]. Further developments of the FPV method was used to study compressible reacting flows by Saghafian et al. [20] by scaling the chemical reaction term of the progress variable equation using the local density and temperature. The results show that their method is able to capture the main features of supersonic flames [20]. However, Saghafian et al. noted that this approach is unable to capture auto-ignition processes due to the limitations of the FPV approach.

In practical high-enthalpy scramjet engines, the air stream is featured by high Mach number and stagnation temperature [22,23]. The reactant temperature before ignition may be higher than the auto-ignition temperature, and the ignition delay time is considerably reduced, comparably to fuel/air mixing time. Therefore, auto-ignition may act as an important mechanism for flame initiation and stabilization in high-enthalpy supersonic flows [24]. Furthermore, under some conditions, the MILD (moderate, intense, or low dilution) conditions [25] can be satisfied theoretically in supersonic combustion [26], i.e. $(\Delta T - T_{ig})/T_u < 0$. Here $\Delta T = T_b - T_u$ is the temperature difference between the burned and unburned gases, whilst $T_{ig}$ is the auto-ignition temperature. Therefore, their intrinsic resemblance provides a novel perspective from MILD combustion for modelling turbulent supersonic combustion with high-enthalpy oxidants.

In this work, the tabulated chemistry method based on Perfectly Stirred Reactor (PSR) model [17] is developed for LES (abbreviated as "LES−PSR" hereafter) of supersonic combustion, and temperature dependence and density correction modification for source term of the progress variable are considered, to account for the effects of viscous heating and compressibility on supersonic flame dynamics. *A prior* analysis of the forgoing effects and then LES of an auto-igniting hydrogen jet flame in supersonic vitiated co-flowing jet [27] are conducted. The objective is to assess the LES−PSR model in predicting the reactive scalar evolutions and auto-ignition dynamics in supersonic combustion. The rest of the paper is structured as follows. Section 2 describes the governing equations and combustion model, while Section 3 performs *a priori* analysis of compressibility effects. Section 4 gives the supersonic jet flame information and numerical implementation. The results are presented and discussed in Section 5 and the main conclusions are summarized in the final section.

## 2. Governing equation and combustion model

## 2.1 *LES governing equation*

The Favre-filtered equations of mass, momentum and energy conservation for turbulent compressible reacting flows are solved in this work

$$\frac{\partial \bar{\rho}}{\partial t} + \nabla \cdot [\bar{\rho}\tilde{\mathbf{u}}] = 0, \tag{1}$$

$$\frac{\partial(\bar{\rho}\tilde{\mathbf{u}})}{\partial t} + \nabla \cdot [\tilde{\mathbf{u}}(\bar{\rho}\tilde{\mathbf{u}})] + \nabla \bar{p} + \nabla \cdot (\bar{\mathbf{T}} - \mathbf{T}^{sgs}) = 0, \tag{2}$$

$$\frac{\partial(\bar{\rho}\tilde{E})}{\partial t} + \nabla \cdot [\tilde{\mathbf{u}}(\bar{\rho}\tilde{E})] + \nabla \cdot [\tilde{\mathbf{u}}\bar{p}] + \nabla \cdot (\bar{\mathbf{T}} \cdot \tilde{\mathbf{u}}) + \nabla \cdot (\bar{\mathbf{J}} - \mathbf{E}^{sgs}) = 0, \tag{3}$$

where $t$ is time, and $\nabla \cdot (\cdot)$ is the divergence operator. The operators $\overline{(\cdot)}$ and $\widetilde{(\cdot)}$ denote conventional and Favre filtering respectively. $\bar{\rho}$ and $\bar{p}$ are respectively the filtered density and pressure, whereas $\tilde{\mathbf{u}}$ is the Favre-filtered velocity vector. $\bar{\mathbf{T}} = -2\mu \text{dev}(\bar{\mathbf{D}})$ is the filtered viscous stress tensor. The dynamic viscosity $\mu$ is temperature dependent and is predicted with Sutherland's law. $\bar{\mathbf{D}} \equiv [\nabla\tilde{\mathbf{u}} + (\nabla\tilde{\mathbf{u}})^T]/2$ is the deformation tensor and its deviatoric component, i.e. $\text{dev}(\bar{\mathbf{D}})$, is defined as $\text{dev}(\bar{\mathbf{D}}) \equiv \bar{\mathbf{D}} - \text{tr}(\bar{\mathbf{D}})\mathbf{I}/3$ with $\mathbf{I}$ being the unit tensor. The filtered total energy $\tilde{E}$ is defined as $\tilde{E} = \tilde{h} + |\tilde{\mathbf{u}}|^2/2 + k_{sgs}$ with $\tilde{h}$ being the absolute enthalpy and $k_{sgs}$ being the sub-grid scale kinetic energy. $\bar{\mathbf{J}} = -\lambda\nabla\tilde{T}$ in Eq. (3) is the filtered diffusive heat flux, where $\tilde{T}$ is the filtered temperature. Note that heat transfer due to mass transfer (e.g. different heat contents of various species or Dufour effect) is not considered in this work. $\lambda$ is the molecular thermal conductivity, and estimated using the Eucken approximation [28], i.e. $\lambda = \mu C_v(1.32 + 1.37 \cdot R/C_v)$, where $C_v$ is the heat capacity at constant volume and obtained using $C_v = C_p - R$. Here $C_p = \sum_{m=1}^{M} \tilde{Y}_m C_{p,m}$ is the heat capacity at constant pressure, and $C_{p,m}$ is estimated using JANAF polynomials [29]. $M$ is the total number of species and $\tilde{Y}_m$ is the filtered mass fraction of *m*-th species. $R$ is specific gas constant and is calculated from $R = R_u \sum_{m=1}^{M} Y_m MW_m^{-1}$. $MW_m$ is the molar mass of *m*-th species and $R_u$ is universal gas constant. The filtered pressure $\bar{p}$ is calculated from the ideal gas equation of state, i.e.

$$\bar{p} = \bar{\rho} R \tilde{T}. \tag{4}$$

The sub-grid scale stress tensor $\mathbf{T}^{sgs}$ in Eq. (2) reads

$$\mathbf{T}^{sgs} = \bar{\rho}(\widetilde{\mathbf{u}\mathbf{u}} - \tilde{\mathbf{u}}\tilde{\mathbf{u}}) = \frac{2\bar{\rho}k_{sgs}\mathbf{I}}{3} - 2\mu_{sgs}\text{dev}(\overline{\mathbf{D}}), \tag{5}$$

where $k_{sgs}$ and $\mu_{sgs}$ are the sub-grid scale kinetic energy and viscosity, respectively. In the present work, they are closed using the constant Smagorinsky model [30,31]. Therefore, $k_{sgs}$ and $\mu_t$ are respectively modelled as

$$k_{sgs} = \left(\frac{-b + \sqrt{b^2 + 4ac}}{2a}\right)^2 \text{ and } \mu_t = C_k \bar{\rho} \Delta \sqrt{k_{sgs}}, \tag{6}$$

where $a = C_\varepsilon/\Delta$, $b = 2\text{tr}(\overline{\mathbf{D}})/3$ and $c = -2C_k\Delta\text{dev}(\overline{\mathbf{D}}):\overline{\mathbf{D}}$ are the coefficients. $C_k$ and $C_\varepsilon$ are constants and take the values of 0.094 and 1.048, respectively [31]. $\Delta$ is the LES filter size and is taken from cubic root of an LES cell volume $V_{cell}$, i.e. $\Delta = V_{cell}^{1/3}$. The sub-grid scale energy flux $\mathbf{E}^{sgs}$ in Eq. (3) takes the following form

$$\mathbf{E}^{sgs} = -\frac{\mu_t}{Pr_t} C_p \nabla \tilde{T} - (\mu + \mu_{sgs})\nabla k_{sgs} + \tilde{\mathbf{u}}\,\mathbf{T}^{sgs}, \tag{7}$$

where the turbulent Prandtl number $Pr_t$ is set as 0.7 [32].

## 2.2 Combustion model

In supersonic combustion, there are two main flow-chemistry interactions at play – subgrid mixing/combustion, and flow compressibility induced mixture reactivity/thermochemical state changes. To account for these effects, we employ the highly efficient and yet reasonably accurate tabulated chemistry approach with carefully chosen lookup table control parameters. Ideally, one could form a six-dimensional table parameterized by mixture fraction, $\tilde{Z}$, its variance, $\widetilde{Z''^2}$, a progress variable, $\tilde{c}$, its variance, $\widetilde{c''^2}$, absolute enthalpy, $\tilde{h}$, and pressure, $\bar{p}$ to include all of these effects. However, such an exercise would not fit with practical interests for a number of reasons, e.g., large memory requirement for loading the look-up table. In order to reduce the table dimension, we performed several trail and test runs to find the sensitivities of the LES results to these parameters. In short, in addition to the mixture fraction, $\tilde{Z}$, and progress variable, $\tilde{c}$, it is found that mixture fraction variance, $\widetilde{Z''^2}$, was important for all the grids tested. This is not surprising since the small scale SGS mixing significantly influences the autoignition behavior in supersonic combustion. The progress variable variance, $\widetilde{c''^2}$, on the other hand, was observed to be rather small due to the small mesh resolution used [20] (see Section 4.2) and hence it is neglected for the present study. For $\tilde{h}$ and $\bar{p}$, from a modelling perspective it is rather straightforward to choose to keep $\tilde{h}$, because the initial temperature clearly influences the ignition delay (in terms of orders of magnitude) as well as the flammability range. By contrast, the pressure effects on the species and reaction rate profiles appear as a scaling effect in the progress variable space for a given mixture fraction [20], which has motivated us to use the density ratio with a power-law scaling for the reaction rate modelling (see Eq. 11).

Therefore, further to the above equations in Section 2.1, the Favre-filtered equations for mixture fraction, $\tilde{Z}$, its variance, $\widetilde{Z''^2}$, and a progress variable, $\tilde{c}$, are solved to model scalar mixing and partially premixed combustion [33,34]. The mixture fraction is defined as the mass fraction of the composition from the fuel stream in the local mixture. The normalized progress variable is defined as $c \equiv \psi/\psi^{Eq}$, where $\psi = Y_{H_2O}$ and $\psi^{Eq}$ is the equilibrium H$_2$O mass fraction for the local mixture [34]. Their governing equations respectively read [35]

$$\bar{\rho}\frac{D\tilde{Z}}{Dt} = \nabla \cdot [(\mathcal{D} + \mathcal{D}_t)\nabla\tilde{Z}], \tag{8}$$

$$\bar{\rho}\frac{D\widetilde{Z''^2}}{Dt} = \nabla \cdot \left[(\mathcal{D} + \mathcal{D}_t)\nabla\widetilde{Z''^2}\right] + 2\mathcal{D}_t|\nabla\tilde{Z}|^2 - 2\bar{\rho}\tilde{\chi}_{Z,\text{sgs}}, \tag{9}$$

$$\bar{\rho}\frac{D\tilde{c}}{Dt} = \nabla \cdot [(\mathcal{D} + \mathcal{D}_t)\nabla\tilde{c}] + \bar{\dot{\omega}}_c^*, \tag{10}$$

in which $D(\cdot)/Dt$ is the substantial derivative. The molecular mass diffusivity $\mathcal{D}$ is calculated through $\mathcal{D} = \lambda/\rho C_p$ with unity Lewis number assumption, and $\lambda$ is estimated through Eucken approximation [28]. In high-speed flows, strong convection and compressibility dominate molecular transport [2], e.g. mass diffusivity and thermal conductivity. Therefore, the Eucken approximation and unity-Lewis number assumption are used. The SGS eddy diffusivity $\mathcal{D}_t$ is estimated from $\mathcal{D}_t = \mu_t/Sc_t$ with turbulent Schmidt number $Sc_t = 0.7$ [4]. The dissipation rate of the mixture fraction variance is modeled as $\tilde{\chi}_{Z,sgs} = C_Z(\mu_t/\Delta^2)\widetilde{Z''^2}$ with $C_Z = 2.0$ [36]. The modelling method for the source term $\bar{\dot{\omega}}_C^*$ for $\tilde{c}$ equation is detailed in the following part.

Zero-dimensional unsteady adiabatic PSR equations with two inflow streams are solved for a range of initial conditions, which has been used for modelling MILD combustion by Chen et al. [17]. Here one should note that at the fundamental level, the combustion occurs and proceeds only if there is fuel and oxidizer at correct proportions (within the flammability limits) and at correct temperature. Hence, the subgrid volume can be seen as a reactor irrespective of premixed or non-premixed combustion mode and the subgrid level diffusion may be negligibly small if the residual scalar and velocity fluctuations are negligible. This is ensured through correct numerical grid resolution guided by Pope's criterion [37]. Furthermore, supersonic combustion can be seen as MILD combustion through timescale analysis as shown in Ref. [26] and it has been shown that PSR model is a suitable model for turbulent MILD combustion at subgrid scales through DNS analysis [38]. In addition to applicable for premixed combustion by definition, the weak (so-called) non-premixed combustion in supersonic combustion can also be seen a collection of reactors at appropriate states depending on the local mixture fraction, progress variable and their variances. Indeed, the reactive structure in phase-space (mixture fraction and progress variable space) was shown to be very similar for premixed, non-premixed and reactor canonical models by Doan et al [39,40]. Hence, the non-premixed combustion related structures can also be captured using the PSR model.

In the PSR model for low Mach number flows (Ma < 0.3) [17], the effects of compressive heating and pressure jump across shock fronts are not considered. However, in supersonic flows (Ma > 1), these effects become significant and should be considered in the PSR tabulation. Therefore, in this work, it is extended to account for the effects of various physical processes involved in supersonic combustion, including mixture stratification, compressive heating and sudden pressure jump at shock fronts. The PSR equations are solved for varying mixture fraction, initial temperature and background pressure. The viscous heating and shock waves in supersonic flows increase the static temperature of the mixture and their effects on combustion can be represented correctly. Generally, the ignition delay time decreases with increased mixture temperature and the heat release rate magnitude increases significantly with increased pressure [20]. Thus, the effects of pressure and temperature variations on the reactions should be considered as additional dimensions to the thermo-chemical table.

Saghafian et al. [20] use a scaling with involving two exponents, one for density ratio and another one for temperature ratio. The auto-ignition delay time is known to be very sensitive to mixture temperature. The temperature effect due to the compressive heating in supersonic combustion is taken into account by incorporating the PSR simulations with varying initial temperatures and using the thermo-chemical enthalpy as a look-up control parameter. The magnitude of the heat release rate increases with pressure but its shape in $c$ space does not change significantly in the PSR. However, this shape is found to be sensitive to the initial mixture temperature. Thus, to include the pressure effects due to the shock wave, the reaction rate source term in Eq. (10) is modeled using a scaling for density ratio as

$$\frac{\overline{\dot{\omega}}_c^*}{\overline{\dot{\omega}}_{c_0}} = \left(\frac{\bar{\rho}}{\bar{\rho}_0}\right)^{\tilde{\alpha}_\rho}, \qquad (11)$$

where the subscript "0" refers to the variables from atmospheric pressure PSR calculations. The influence of temperature is included explicitly by using the absolute enthalpy $\tilde{h}$ (i.e., chemical and sensible enthalpies) as one of the control variables in the look-up table, which is deduced on the fly using the transported total energy in Eq. (3). The exponent $\tilde{\alpha}_\rho$ is evaluated as per the above scaling at peak heat release point in the PSR model with different pressures (the peak locations are more or less the same in normalized progress variable space) in the reactor, and then the mean is taken for each mixture fraction and enthalpy. The fluctuation of $\tilde{\alpha}_\rho$ with different pressures does not exceed 5% at worst scenarios observed in the calculations. It is also to be noted that the flammability limits vary considerably with both pressure and initial temperature, particularly at the rich side, and thus the averaging is taken only from the cases with non-zero reaction rates.

Hence, the value of this exponent in the look-up table is obtained using $\tilde{\alpha}_\rho = \iint \alpha_\rho P(Z,h) dZ dh$, where $P(Z,h) = P_\beta(Z) P_\delta(h)$ is the joint PDF modelled using a beta and a delta function for $Z$ and $h$, respectively. $\overline{\dot{\omega}_{C_0}}$ and $\bar{\rho}_0$ and other thermo-chemical quantities, in a vector form $\underline{\tilde{\phi}}$, are obtained using $\underline{\tilde{\phi}} = \iiint \alpha_\rho P(Z,h,c) \, dZ \, dh \, dc$ with $P(Z,h,c) = P_\beta(Z) P_\delta(h) P_\delta(c)$.

## 3 *A priori* analysis of compressibility effects

A *priori* analysis of the combustion model in modelling compressibility effects is presented in this Section. Figure 1 shows mass fractions of selected species, including $H_2$, $O_2$, OH, H, $HO_2$, and source term of the progress variable equation versus reaction progress variable at different initial temperatures, i.e., 1,250, 1,300, 1,350 and 1,400 K. The results in Fig. 1 are calculated with the same pressure and mixture fraction, i.e., $p = 1.0$ atm and $Z = 0.03$. Figure 2 shows the counterpart results at different pressures, e.g., 0.5, 1.0, 1.5 and 2.0 atm. The results in Fig. 2 are calculated with the same initial pressure and stoichiometric mixture fraction, i.e., $T = 1,250$ K and $Z = 0.03$. Major species such as $H_2$ and $O_2$ mass fractions change little with varying initial reactant temperature or pressure. The minor species such as OH, H and $HO_2$ are slightly affected by the initial temperature and pressure. However, the contributions of these minor species to the mixture properties are much smaller than those of the major species. Moreover, the source term of the progress variable versus reaction progress variable is significantly affected by the initial temperature and pressure. An important observation here by comparing Figs. 1f and 2f is that the peak in the reaction rate profile shifts towards smaller $c$ values as initial temperature increases, e.g., from about $c$ = 0.5 for 1,250 K to 0.4 for 1,400K, while the peak position is almost the same when pressure changes from 0.5 to 2 bar. This justifies our modelling strategy described in Section 2.2, which is to use initial temperature as a tabulation control variable and apply a scaling equation to account for the pressure effect on the reaction rate. Note that the temperature and pressure ranges considered include most auto-ignition temperature and pressure for hydrogen combustion in the supersonic combustion conditions [1].

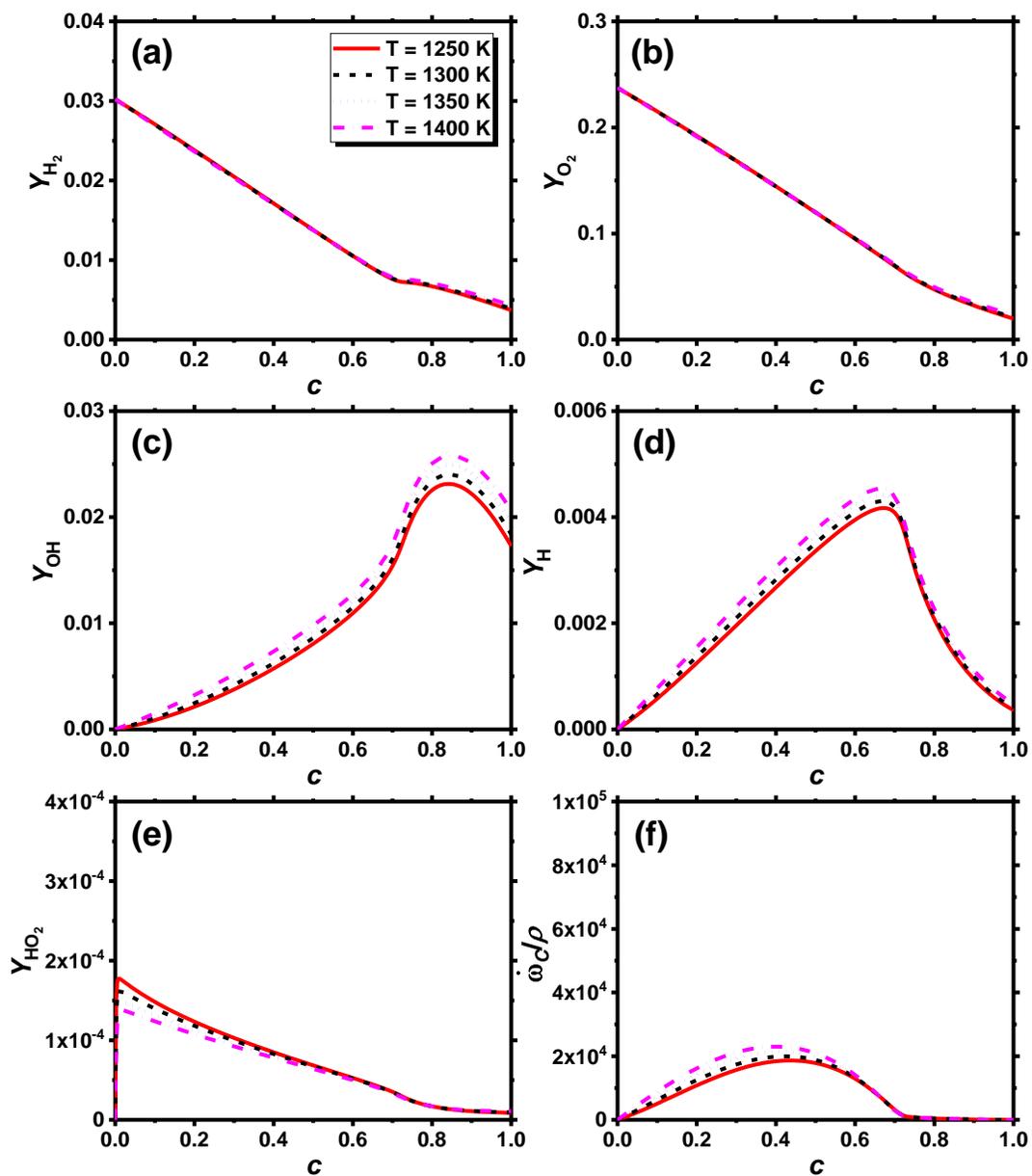

Fig. 1 Mass fractions of $H_2$, $O_2$, OH, H, $HO_2$ and source term of progress variable versus reaction progress variable at different temperatures. The pressure is 1 atm and mixture fraction is 0.03.

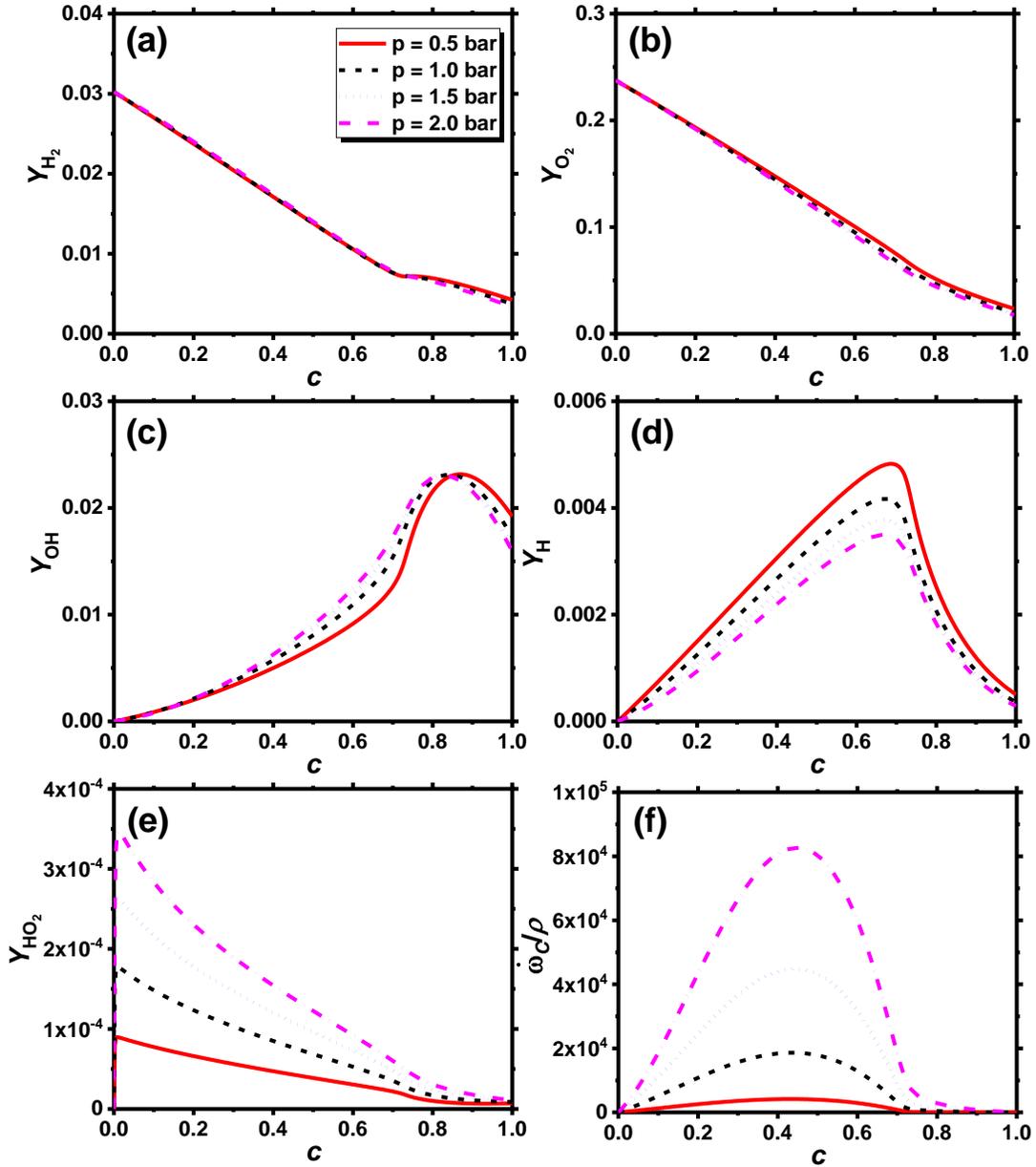

Fig. 2 Mass fractions of $H_2$, $O_2$, OH, H, $HO_2$ and source term of progress variable versus reaction progress variable at different pressures. The temperature is 1,250 K and mixture fraction is 0.03.

The thermo-chemical quantities in the look-up table are generated at the reference pressure, i.e., $p$ = 1 atm. For the influence of temperature, the absolute enthalpy $\tilde{h}$ is used as one of the control variables for the look-up table [17]. For the influence of pressure, the scaling in Eq. (11) for the source term is used. Note that the exponent $\tilde{\alpha}_\rho$ in Eq. (11) is a function of mixture fraction and the absolute enthalpy in this combustion model. Figure 3 shows the exponent $\tilde{\alpha}_\rho$ versus mixture fraction with different temperatures, e.g., 1,250 K, 1,300 K, 1,350 K and 1,400 K. It is shown that $\tilde{\alpha}_\rho$ gradually increases and then rapidly decreases to zero with increased mixture fraction for a given temperature. The mixture fraction at which the sudden drop occurs essentially correspond to the rich flammability limit, which increases significantly with the PSR initial temperature as one would expect. This discontinuity of $\tilde{\alpha}_\rho$ in mixture fraction space is non-problematic because the base reaction, $\bar{\dot{\omega}}_{c_0}$, in Eq. (11) tends to zero naturally as mixture fraction approaches the rich limit. In order to confirm the validity of Eq. (11) to capture the compressibility effects, comparisons between the source term under an elevated pressure and rescaled source term from the background pressure is presented in Fig. 4. In addition, a constant exponent $\tilde{\alpha}_\rho = 2$ as suggested by Pecnik et al. [18] is also used here for comparison in Fig. 4. It is shown that much lower reaction source term is obtained with the same correction as Eq. (11) by using the constant exponent. The results suggest that the reaction source term at $p = 1.5, 2$ and 5 atm are well reproduced through using variable exponents $\tilde{\alpha}_\rho$ for the density ratio as indicated Eq. (11). It is noted that the reaction rate profile has a noticeable shift between the PSR results and the scaled profile using Eq. (11). This is because at higher pressure, the peak reaction rate occurs at a slightly larger progress variable value (i.e., at a larger $H_2O$ mass fraction). This is not problematic for the present test case because: i) this shift is only about 2-3% in the $c$ space; and ii) the local pressure jump due to shock presence is well below 5 atm. Therefore, we apply a single exponent to describe the pressure dependence for all the pressures considered.

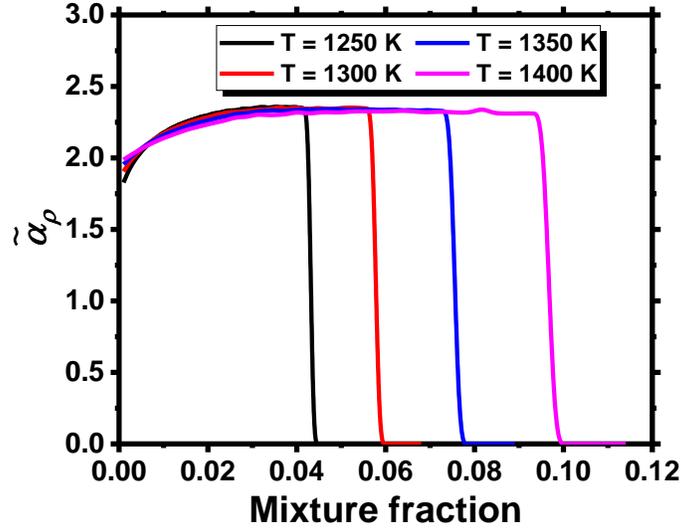

Fig. 3 The exponent $\tilde{\alpha}_\rho$ versus mixture fraction with different temperatures.

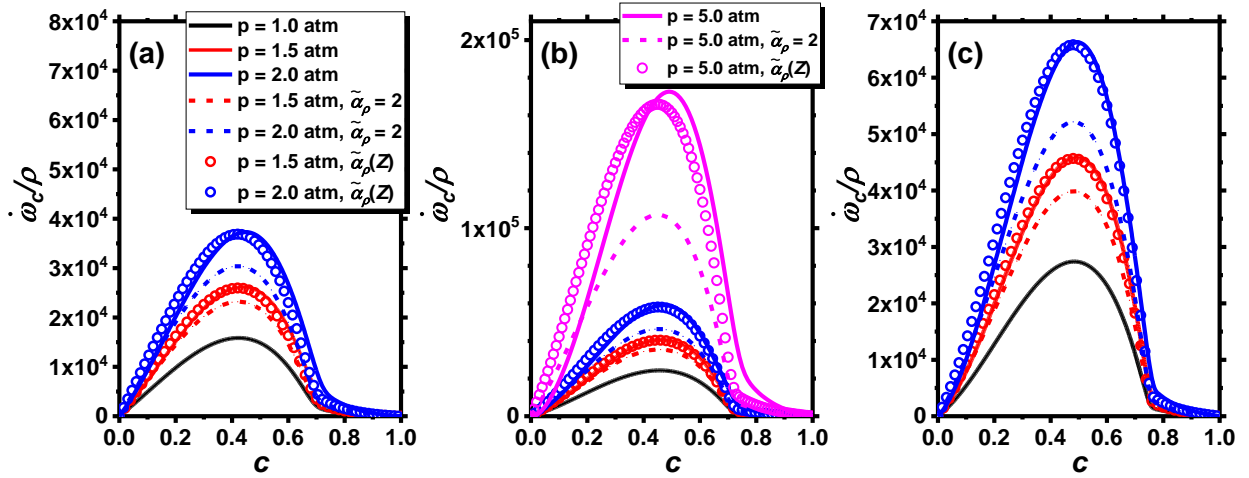

Fig. 4 Source term of progress variable equation versus reaction progress variable for different mixture fractions: (a) 0.02, (b) 0.03 and (c) 0.035. The temperature is 1,250 K.

## 4. Flame information and numerical implementation

### 4.1 *Cheng supersonic flame*

A Mach 2 supersonic lifted hydrogen jet flame has been investigated by Jarret et al. [41] from NASA Langley Research Center and Cheng et al. [27] from Vanderbilt University. Hereafter it is called as Cheng supersonic flame. The schematic of the burner and a long exposure visual photo of the stably burning flame are shown in Figs. 5(a) and 5(b), respectively. Sonic hydrogen injected into a co-flowing supersonic stream is performed to study auto-ignition and stabilization of combustion in supersonic streams. The fuel and oxidizer conditions are listed in Table 1. The fuel is 100% hydrogen, whereas the oxidizer stream is composed of 20.1% $O_2$, 54.4% $N_2$ and 25.5% $H_2O$ by volume. The inner diameter of the cylindrical central fuel jet is $D = 2.36$ mm, whereas the inner and outer diameters of the concentric annular co-flowing jet are 3.81 and 17.78 mm respectively. The fuel velocity is 1,780 m/s with a static temperature of 545 K (the corresponding stagnation temperature is 1,750 K). The vitiated air stream is the combustion product of hydrogen and $O_2$-enriched air, which is accelerated through an axi-symmetric convergent-divergent nozzle and reaches Mach 2 at the exit. Hydrogen auto-ignition occurs downstream of the burner exit due to the high co-flowing temperature as well as the efficient reactant mixing facilitated by the strong shearing between the sonic fuel jet and surrounding Mach 2 vitiated co-flowing. Therefore, the flame stabilizes at a distance downstream of the fuel jet and the measured flame lift-off height is $25D$ based on the flame photo in Fig. 5(b). More detailed information about this burner can be found in Refs. [27,42].

Table 1. Fuel and co-flowing conditions.

| *Parameter* | *Fuel jet* | *Co-flowing jet* |
|---|---|---|
| Pressure (Pa) | 112,000 | 107,000 |
| Temperature (K) | 545 | 1,250 |
| Mach number | 1.0 | 2.0 |
| Velocity (m/s) | 1,780 | 1,420 |
| $H_2$ mole fraction | 1.0 | 0.0 |
| $O_2$ mole fraction | | 0.201 |
| $N_2$ mole fraction | 0.0 | 0.544 |
| $H_2O$ mole fraction | | 0.255 |

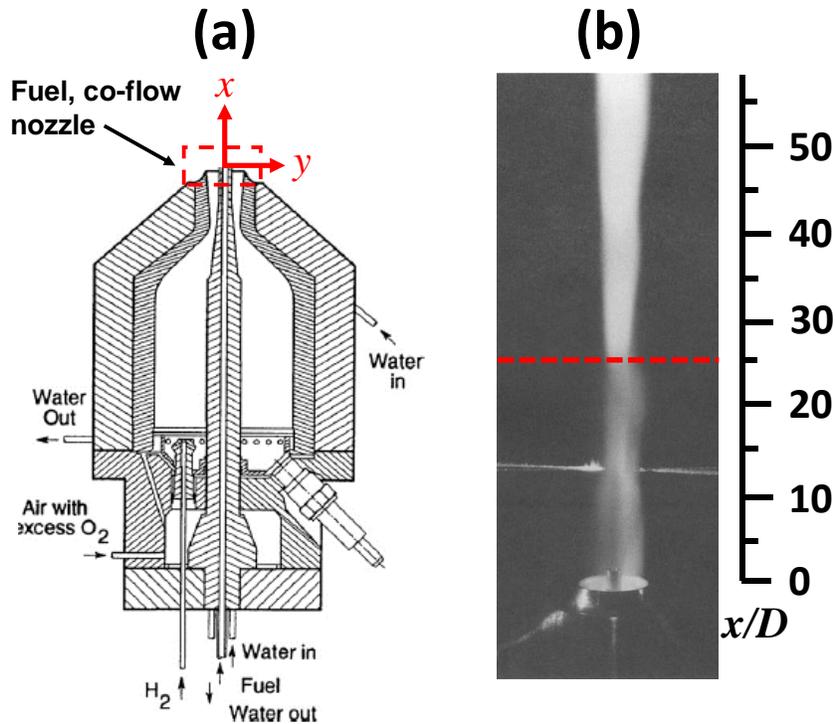

Fig. 5 (a) Schematic of the supersonic burner and (b) long exposure visual photo of the supersonic flame [27]. Dashed line in (b) denotes the measured lift-off height (25$D$, $D$ = 2.36 mm is the jet diameter).

The Coherent Anti-Stokes Raman Scattering (CARS) measurements are used to measure the mean temperature as well as the species concentration, whilst the Laser Doppler Anemometry (LDA) is to obtain the mean velocity distributions [41]. Based on 500 or 2000 independent laser shots, Cheng et al. further obtain the cross-sectional profiles of mean and RMS of the major species mole fractions (including $O_2$, $H_2$, $H_2O$, $N_2$ and OH), as well as temperature at seven streamwise stations. The scatter plots of the temperature and foregoing major species mole fractions are also available at six pointwise locations [27]. Furthermore, Using the velocity measurements by Jarrett et al. [41] and Raman measurements by themselves, Cheng et al. also estimate the fluid mechanical scales, including Kolmogorov length / time scales, as well as the integral length scale. It is also shown that the turbulent fluctuating velocities are about 5%−10% of the local velocities [41], whilst the species and temperature fluctuations are as high as 40% and 20%, respectively [27]. Therefore, this lifted flame is characterized by strong combustion unsteadiness due to high fluctuations of velocity and reactive scalars. These detailed experimental measurements and combustion / flow field analysis made in Refs. [27,41] offer us an ideal but challenging test case to validate the PSR−LES model in handling the interactions between turbulence, flow discontinuities, reacting mixing and chemical reactions.

## 4.2 *Simulation details*

A cylindrical domain of $100D \times 60D \times 2\pi$ (axial, radial and azimuthal directions, respectively) is used for LES selected here. The coordinate origin is at the center of the fuel jet exit with $x$ and $y$ being the streamwise and radial coordinates respectively, as indicated in Fig. 5(a). The inlet plane of the LES domain is at 1.18 mm upstream of the burner exit and thus a part of the fuel pipe and co-flowing nozzle are included in the computational domain, enclosed in Fig. 5(a) by the red dashed box. Note that the whole geometry of the nozzle configuration is included in the LES of the same burner by Moule et al. [4], and the predicted flow and flame features of interest (e.g. shock structure, flame lift-off height and reactive scalar profiles) generally bear close resemblance to the results of LES without detailed nozzle configuration [3,9,43]. Therefore, our LES domain is expected to be sufficient to accurately reproduce the major aerodynamic and combustion characteristics of this supersonic flame.

This domain is discretized with about 26 million hexahedrons and it is refined to have a cell size of about 0.16 mm in a central region of size $44D \times 5D \times 2\pi$ so that the shear flows, scalar mixing and unsteady auto-ignition processes in induction ($x < 20D$) and stabilization ($x < 40D$) zones can be well captured [4]. This cell size is around 8−16 Kolmogorov scale (0.01-0.02 mm, estimated from [27]) and is comparable to those used by Boivin et al. with 0.1−0.4 mm [43], Bouheraoua et al. [3] with 0.06−0.24 mm and Moule et al. [4] with 0.1−0.2 mm, which are tabulated in Table 2. Despite the fine resolution, there are still some unresolved turbulence and conserved scalar fluctuations that require the sub-grid model (see Figs. S1 and S2 in the Supplemental Material). Also, the necessity of the existing fine cells has also been confirmed by comparing the results with a coarse mesh (see Figs. S3 and S4 in the Supplemental Material).

*A posteriori* analysis of the present LES mesh resolution is made in terms of the ratio of the SGS to molecular viscosities, $\mu_{sgs}/\mu$, in the central region of $44D \times 5D \times 2\pi$ (from the jet exit in *x*-direction), where the reactant mixing and combustion proceeds. Figure 6(a) shows this ratio versus the filtered heat release rate, and one can see that for most of the locations with significant heat release rate ($>1\times10^9$ J/m$^3$/s) and pronounced temperature rise ($>1300$ K), $\mu_{sgs}/\mu$ is predominantly below 1.0, indicating that the LES is well resolved in the combusting regions. Furthermore, the power spectral density analysis of axial velocity in the jet shear layer also verifies that the existing resolution lies in the inertial sub-range of turbulence scales (see Fig. S5 in the Supplemental Material).

Moreover, the test incorporating the equation of SGS variance of $\tilde{c}$ [35] into the existing governing equations in Section 2.1 is also performed. Here, the SGS variance of $\tilde{c}$ is solved from [35]

$$\bar{\rho}\frac{D\widetilde{c''^2}}{Dt} = \nabla \cdot \left[(\mathcal{D} + \mathcal{D}_t)\nabla \widetilde{c''^2}\right] + 2\mathcal{D}_t|\nabla\tilde{c}|^2 - 2\bar{\rho}\tilde{\chi}_{Z,\text{sgs}} + 2(\overline{c\dot{\omega}_c^*} - \tilde{c}\bar{\dot{\omega}}_c^*), \qquad (12)$$

where $\widetilde{c''^2}$ denotes the SGS variance of $\tilde{c}$. Figure 6(b) shows the scatter plot of SGS $\tilde{c}$ variance versus heat release rate in the central region of size $44D\times5D\times2\pi$ with refined mesh. It is observed that the SGS $\tilde{c}$ variance is predominantly smaller than 0.01. Hence, use of the delta function for $P_\delta(c)$ is justified. This also consolidates our selection of the tabulation variables without $\tilde{c}$ variance as detailed in Section 2.2.

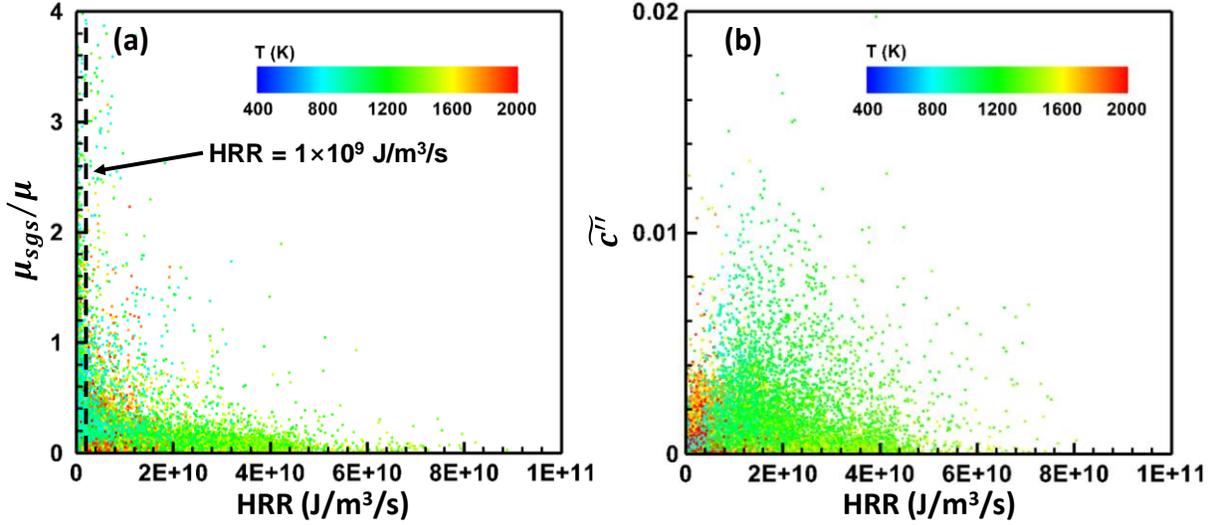

Fig. 6 (a) Scatter plot of $\mu_{sgs}/\mu$ versus heat release rate and (b) Scatter plot of SGS $\tilde{c}$ variance versus heat release rate colored by temperature in the central region of 44$D$×5$D$×2π.

A finite-volume LES solver for multi-component compressible reacting flows *RYrhoCentralFoam* [44] is used, developed from the fully compressible non-reactive flow solver *rhoCentralFoam* [45] in OpenFOAM 5.0. *rhoCentralFoam* employs the KNP method [46] with van Leer limiter to capture shocks and its accuracies has been validated by Greenshields et al. [45] through various non-reacting benchmark tests on the one-dimensional Sod's problem, two-dimensional forward-facing step and supersonic jet flows. Validation of *rhoCentralFoam* in turbulent, high-speed reacting flows with quasi-laminar chemistry model can be found in our recent work on benchmark tests [47], an auto-igniting, cavity stabilized ethylene flame [48] and a coflow hydrogen jet flame [49], and also by others, e.g. Wu et al. [50], Ye et al. [21,51], Piao et al. [52], Cai et al. [53].

The LES equations for momentum and energy (Eqs. 2 and 3) are integrated with an operator-splitting method [45]. Second-order Godunov-type central and upwind-central schemes are used for the convection terms in the Navier-Stokes equations, whereas TVD scheme is used for the convection terms in Eqs. (1)−(3) to bound scalar values. Second-order central difference scheme is applied for the diffusion terms. The second-order implicit backward method is employed for temporal discretization and the time step, $10^{-9}$ s, is chosen to have CFL number < 0.3. Detailed chemistry (19 reactions and 9 species) [54] is used for tabulation, which has been validated against the measured ignition delay at elevated pressures [44].

The boundary conditions of fuel and co-flowing streams are specified, consistent with the experiments listed in Table 1. The inlet turbulence is assumed to be white noise with 5% intensity, since the accurate turbulence flow statistics at both inlets are not measured [27,42]. A synthetic turbulence of 5% of the mean axial velocity is added to the vitiated air injection by Almeida and Navarro-Martinez [9], whilst pre-computed turbulence following a given turbulence spectrum are provided for both fuel and co-flowing jets by Bouheraoua et al. [3]. However, they do not compare the various turbulence specifications, and therefore the effects of the inlet turbulence are not clear. To assess the inlet turbulence effects, an LES with synthetic turbulence inlet is performed. The results (see Figs. A2-A4 in Appendix A) show that both the results with and without the synthetic turbulence show good agreement with the experiments. However, the shock structures would be blurred when the inlet turbulence is included (see Fig. A1). Adiabatic no-slip wall condition is used for the fuel pipe and co-flowing nozzle walls, which is expected to have negligible effects on downstream flame development due to the lift-off characteristics of this flame. Non-reflective condition is used for lateral and outlet boundaries of the computational domain. The mixture fraction is specified to be 1 and 0 for the fuel and co-flowing streams, respectively. The progress variable and mixture fraction variance are set to be zero at the two inlets. In addition to pressure, all the variables are extrapolated with zero-gradient condition at the outlet. A quiescent flow at 1 atm and 298 K is initialized for the LES. A non-reacting flow field is simulated first without $\bar{\bar{\omega}}_c^*$ in Eq. (11), and then combustion is allowed to evolve by including $\bar{\bar{\omega}}_c^*$. The statistics of the combustion case are compiled over 0.4 ms after the effects of initial field is purged (over a period of 0.68 ms). A simulation for physical time of 0.05 ms using 168 cores of ASPIRE 1 Cluster in National

Supercomputing Center Singapore, takes about 24 hours.

Table 2. LES of Cheng supersonic flame [27].

| CFD Solver | Combustion model and chemistry | Mesh resolution | Computational domain | Ref. |
|---|---|---|---|---|
| AVBP [55] | Quasi-laminar chemistry | 6.6 million cells, with minimum cell size 0.1-0.4 mm | Hemisphere domain with 0.85$D$ off burner exit and radius 10,000D | [43] |
| | 5 species, 3 reactions | | | |
| SiTComb [56] | Quasi-laminar chemistry | Three meshes of 4, 32 and 268 million cells, with minimal cells 0.24-0.86 mm, 0.12-0.43 mm and 0.06-0.215 mm | Cylindrical domain (70$D$×30$D$×30$D$) starting in the fuel injection plane | [3] |
| | 5 species, 3 reactions | | | |
| CEDRE [57] | Unsteady PaSR | 31 million cells, with minimal cell size 0.1-0.2 mm | Cylindrical domain, detailed nozzle configuration included | [4] |
| | 9 species, 19 reactions | | | |
| CompReal | Eulerian PDF | 0.2 and 2 million cells | Cylindrical domain (70$D$×60$D$×60$D$) | [9] |
| | 9 species, 19 reactions | | | |
| OpenFOAM [45] | PSR model | 26 million cells, with minimal cell size 0.16 mm | Cylindrical domain (70$D$×60$D$×60$D$) | Current work |
| | 9 species, 19 reactions | | | |

Furthermore, for comparing LES−PSR modelling with other LES of the same flame, different simulation results from Refs. [3,4,9,43] will be included for discussion where necessary in Section 5. The key numerical implementations are briefly summarized in Table 2, including combustion model, chemical mechanism, computational domain, and mesh size.

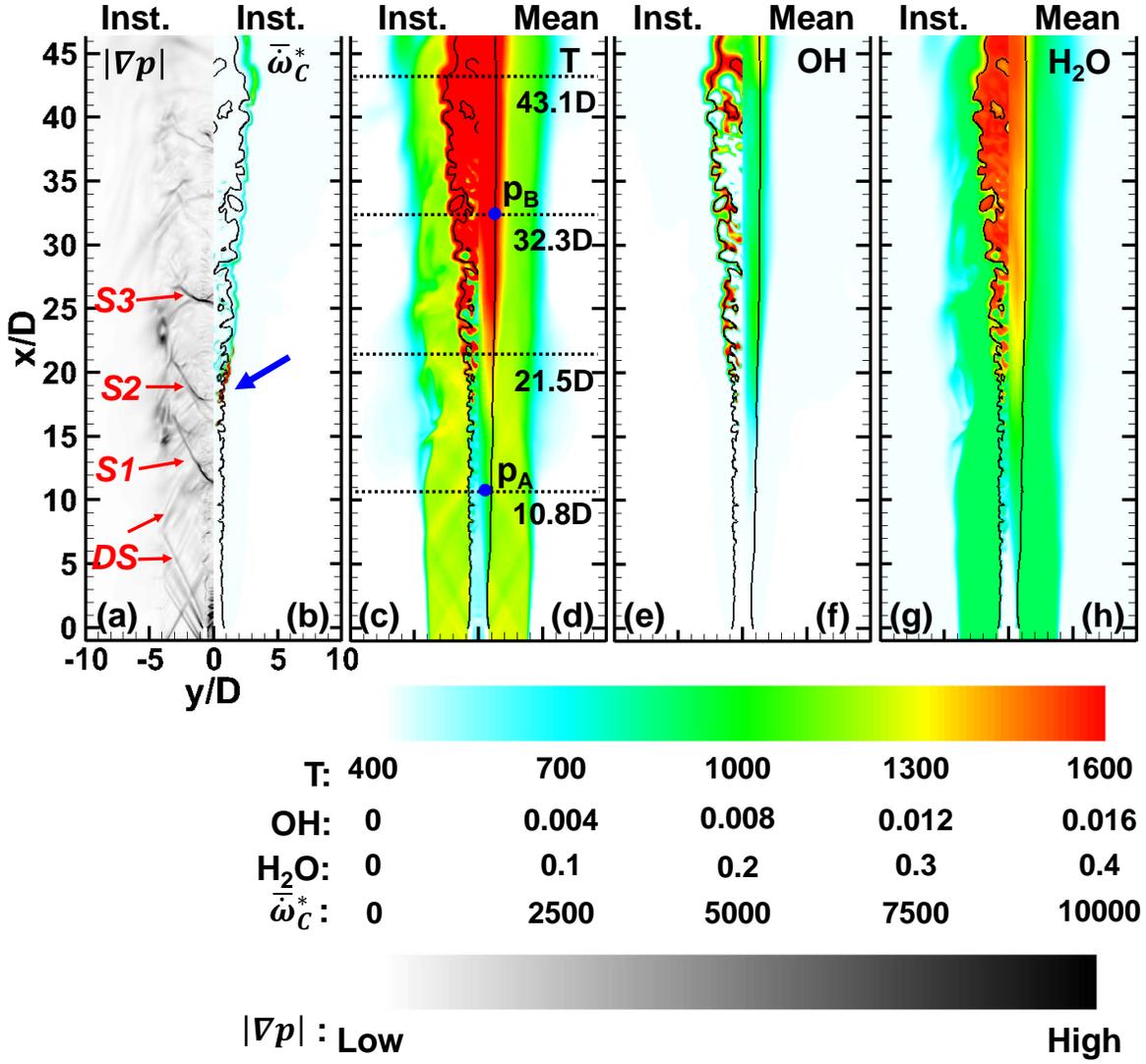

Fig. 7 (a) Pressure gradient magnitude, (b) $\tilde{c}$ equation source term (in 1/s), (c, d) temperature (in K), (e, f) OH mass fraction and (g, h) $H_2O$ mass fraction. Black iso-lines: stoichiometric mixture fraction. "DS" represents the diamond shock and "S1", "S2" and "S3" are V-shaped shocks. $p_A$: probe point A $(x/D, y/D) = (10.8, 0.65)$, $p_B$: probe point B $(x/D, y/D) = (32.3, 1.1)$.

## 5. Results and discussion

### 5.1 *Flow and flame structures*

Figure 7(a) shows the distribution of the flow structure based on the norm of the instantaneous pressure gradient, $|\nabla p|$, which are close to the results captured by Moule et al. [4] with detailed nozzle structure included (see Table 2). The diamond shocks in the co-flowing stream are seen clearly. The first shock (*DS* in Fig. 7a) starts from the jet exit and ends at around 11*D* where a strong V-shaped shock *S1* is formed. Further downstream, the diamond shock becomes unclear, which is probably due to its interactions with local turbulence. However, the V-shaped shocks, *S2* and *S3*, arise at around *x*/*D* = 18 and 25, which play important roles for the formation of auto-ignition spots and flame stabilization. For instance, high $\bar{\dot{\omega}}_C^*$ is observed behind *S2* (see the arrow in Fig. 7b), implying that the pronounced chemical reactions are initiated by pressure and static temperature rise behind *S2*. However, no reaction (low $\bar{\dot{\omega}}_C^*$) behind *S1* is observed, maybe due to the limited mixing there between the hydrogen and co-flowing.

The distributions of temperature, OH and H$_2$O mass fractions are shown in Figs. 7(c)−7(h). Note that the left (right) sub-figures correspond to the respective instantaneous (time averaged) results. Apparently, before *x*/*D* = 20, the temperature is relatively low, which is indicative of limited heat release from chemical reactions and also manifested by low OH and H$_2$O concentrations at these locations. This is qualitatively consistent with the experimental measurements in Fig. 5(b). Nevertheless, around *x*/*D* = 20, pronounced temperature rise (above 2,000 K) can be seen from both instantaneous and mean contours in Figs. 7(c) and 7(d), accompanied by large amount of OH and H$_2$O near the stoichiometric mixture fraction isolines in Figs. 7(e)−7(h). This implies that the chemical reactions are initiated, which corresponds to high $\bar{\dot{\omega}}_C^*$ when *x*/*D* > 20 in Fig. 7(b). The flame base is observed to fluctuate between 10*D* and 25*D* in our simulation, during which the intermittently auto-igniting spots and flame propagation occurs. Similar behaviors were also captured in Refs. [3] and [4], and this will be discussed further in Section 5.4. The predicted mean lift-off height is around 26*D*, determined by the axial distance at which the mean temperature exceeds 1,600 K. Using another criterion (e.g., OH concentration or heat release rate) would yield almost the same results. Our prediction is closer to the measured height of 25*D* [27] (as indicated by the dashed line in Fig. 5b), compared to those (15$D$−35$D$) reported by previous LES studies [3,4,43].

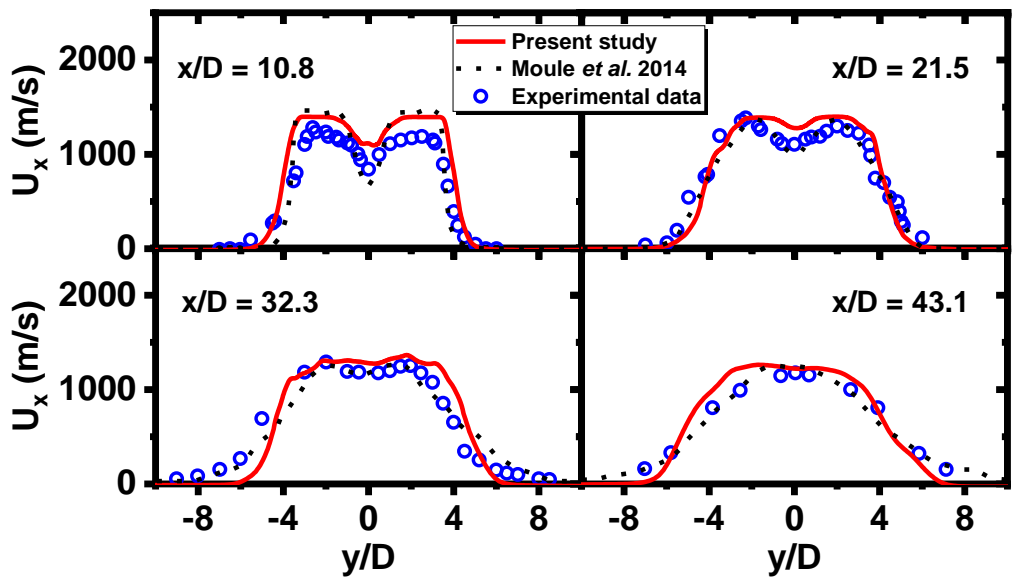

Fig. 8 Radial profiles of mean axial velocity. Comparison with experimental data from Ref. [27] and LES data from Ref. [4].

5.2 *Velocity and scalar statistics in physical space*

Figure 8 shows the radial profiles of mean axial velocity at four streamwise locations in the induction and flame stabilization zones, as marked in Fig. 7. In general, the mean axial velocities at various locations are captured quite well in the present LES−PSR simulation. At *x/D* = 10.8, the velocities in the central jet and surrounding co-flowing stream are slightly over-predicted (by about 13% and 8%, resepctively). This may be because the turbulence used as the boundary conditions are not accurately reproduce the real jet conditions in the experiment. However, such a difference was also reported by Moule et al. [4], who included the detailed nozzle configuration as indicated in Table 2, and in our LES with inlet synthetic turbulence (see Fig. A2 in Appendix A).

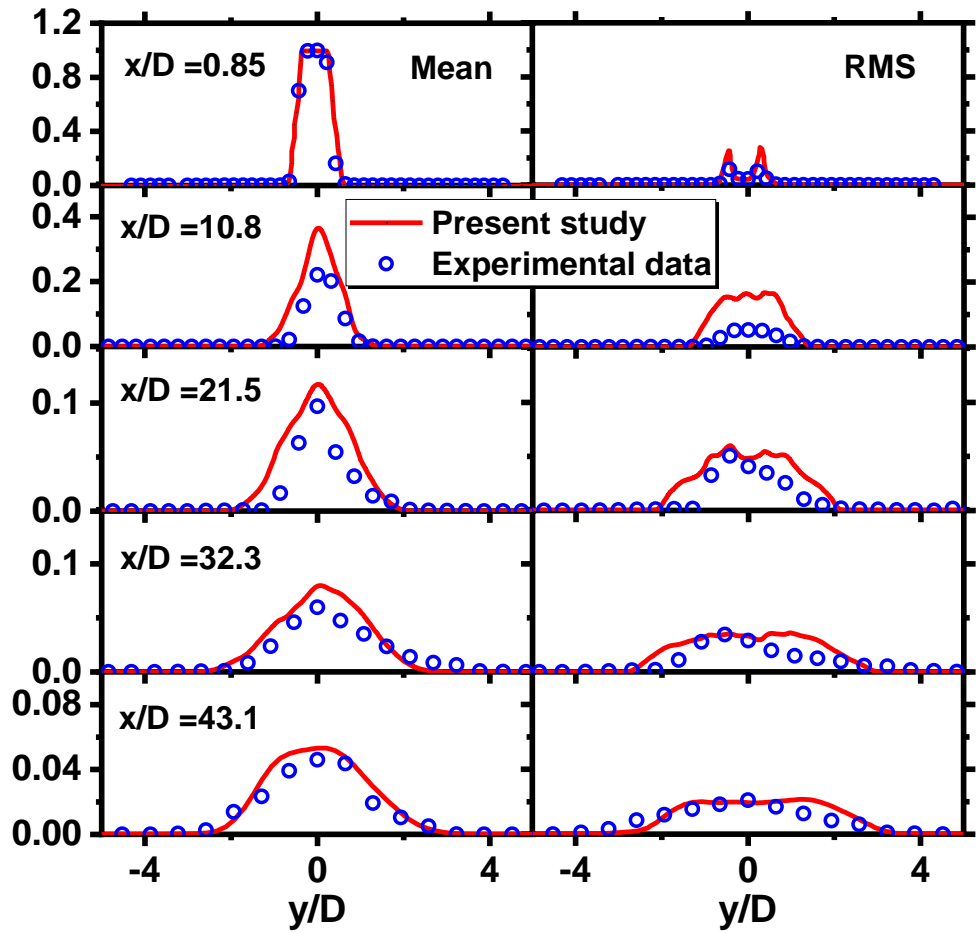

Fig. 9 Radial profiles of mean (left column) and RMS (right column) of mixture fraction. Experimental data from Ref. [27].

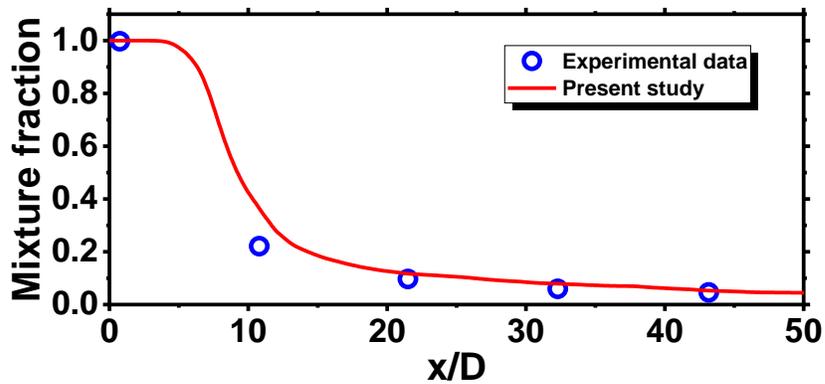

Fig. 10 Centerline profile of mean mixture fraction. Experimental data from Ref. [27].

Figure 9 compares the computed and measured radial variations of mean mixture fraction and its RMS for five streamwise locations, i.e. $x/D$ = 0.85, 10.8, 21.5, 32.3 and 43.1. Note that the RMS in this work is calculated based on the resolved field. At $x/D$ = 0.85 near the jet exit, the mean value of mixture fraction shows fairly good agreement with the experimental data. However, at $x/D$ = 10.8, the mean value along the centerline is overshot by 17.8%, which may be associated with the velocity over-prediction shown in Fig. 8. The higher predicted RMS in the central fuel jet may stem from the stronger turbulence at $x/D$ = 10.8, from the interactions between temproally and spatially evolving shocks, e.g. *DS* and *S1* in Fig. 7(a). Furthermore, in the further downstream locations, our LES results are closer to the measured ones, including the profiles and peak values. In the locations of $x/D$ = 21.5 and 32.3, the measured RMS profiles are not symmstric with respect to $y/D$ = 0. This is caused by the tilted placement of the burner in the expermients, as mentioned in Ref. [27]. Moreover, Fig. 10 shows the centerline profiles of the mean mixture fraction and the predicted mixture fractions show fairly good agreement with the experimental data except for some difference around $x/D$ = 10. Generally, the reactant mixing in this supersonic flow is accurately predicted, which is important for modelling flame auto-ignition and stabilization with the LES−PSR model.

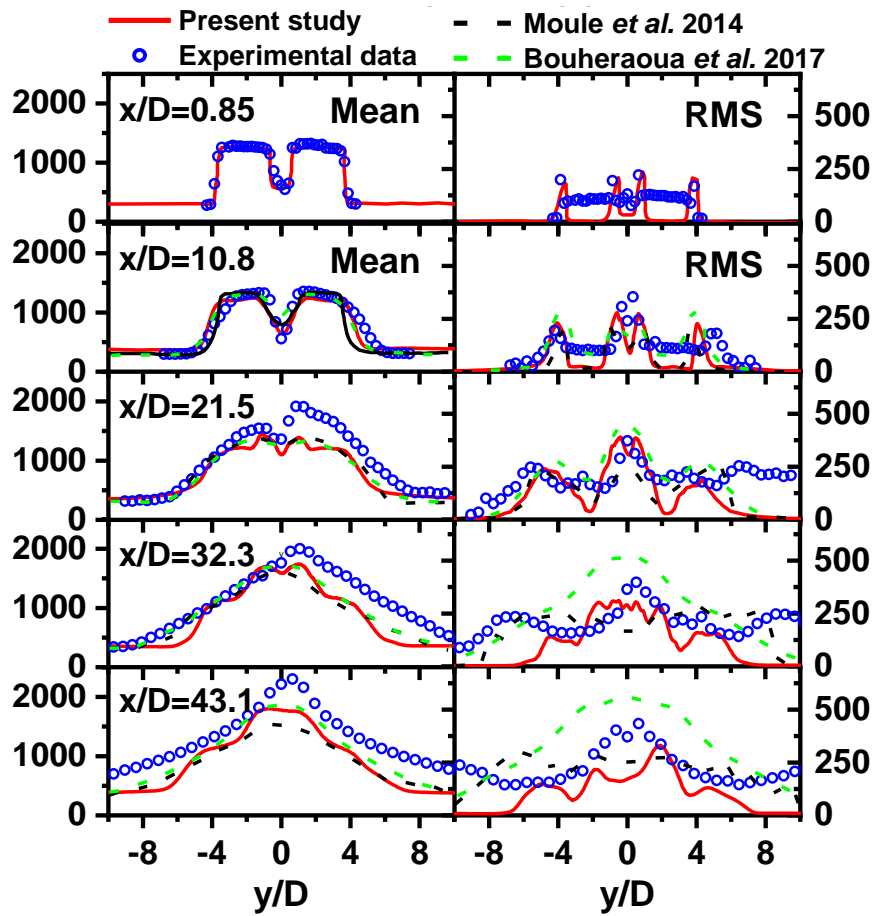

Fig. 11 Radial profiles of mean (left column) and RMS (right column) of temperatures (in K). Experimental data from Ref. [27], whilst the LES data from Refs. [3,4].

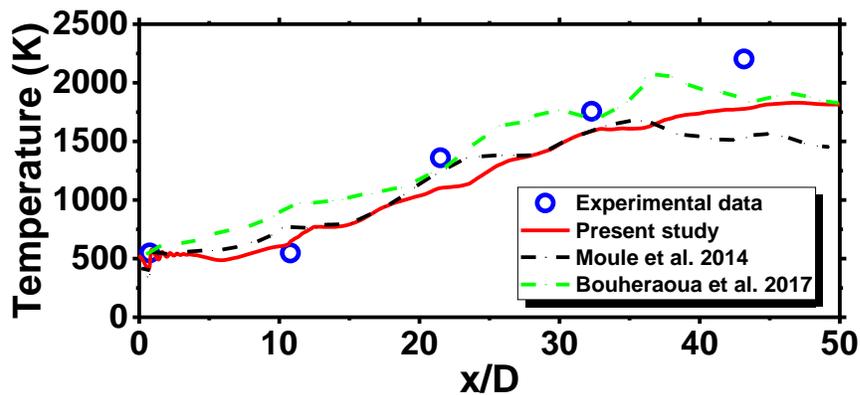

Fig. 12 Centerline profile of mean temperature. Experimental data from Ref. [27], whilst the LES data from Refs. [3,4].

The radial profiles of mean and RMS of temperatures are presented for the same streamwise locations as in Fig. 11. The measured mean temperature profiles are asymmetric due to a small tilt in the burner arrangment [27], and this asymmetry is also observed in Fig. 9 for the mixture fraction RMS. In light of this, we focus on the data for $y < 0$. In general, the computed values agree well with the measurements. At $x/D = 43.1$, the mean temperature is underestimated in our LES, especially in the flame region, $y > -2D$. The mean tempreature from the LES results of Bouheraoua et al. [3] and Moule et al. [4] are also shown in Fig. 11 for comparisons. In general, the differences between the three LES results are small. In the downstream location of $x/D = 43.1$, the centerline mean temperature from our results is closer to that in Ref. [3], which is obtained with a finer mesh (0.06 mm as minimum cell size, as indicated in Table 2). Therefore, one can speculate that predictions of mean temperature in this flame are not sensitive to the sub-grid combustion models.

The distributions of temperature RMS are generally satisfactory for $x/D = 0.85$, 10.8 and 21.5, although the temperature fluctuations inside the co-flowing jet is under-calculated in our work. This discrpancies may be associated with the possible existing chemical reactions, probably from the chemical non-equilibrium in the $H_2/O_2$-enriched combustion product of the co-flowing jet. Moule et al. [4] also under-predict the temperaure fluctuations there, who use unsteady PaSR model and detailed nozzle in their work (see Table 2). On the contrary, the results by Bouheraoua et al. [3] with quasi-laminar chemistry model show better agreement with the measured ones, probably because the finer mesh is used as tabulated in Table 2 and/or the isotropic turbulence specified for the inlets. Furthre downstream after the flame is stabilized, e.g. at $x/D = 32.3$, the temperature RMS inside the flame are well captured, but it is underestimated in the coflowing jet, which is also true for $x/D = 43.1$. This is likely because the mesh resolution used for $|y/D| > 5$ may be insufficient to capture the unsteady behavior of the external mixing layer between the ambient air and co-flowing stream. Different levels of the differences, particularly at the two downstream locations, are also shown from the RMS values predicted by Bouheraoua et al. [3], Moule et al. [4] and Boivin et al. [43], indicating their strong sensitivity to LES numerics. However, these temperature RMS differences in the downstream are expected to have marginal influences on prediction of supersonic flame auto-ignition and stabilization. Moreover, the predicted mean temperature along the centerline is also compared with the experiments [27] and other LES studies [3][4], which is shown in Fig. 12. The mean temperature is well captured in the near field, such as at $x/D$ = 10.8, but underestimated in the downstream. Although the standard Smagorinsky model and similar spatial resolutions (see Table 2) are used in our LES and others, they have different computational domain and different numerical methods, thus probably leading to these differences.

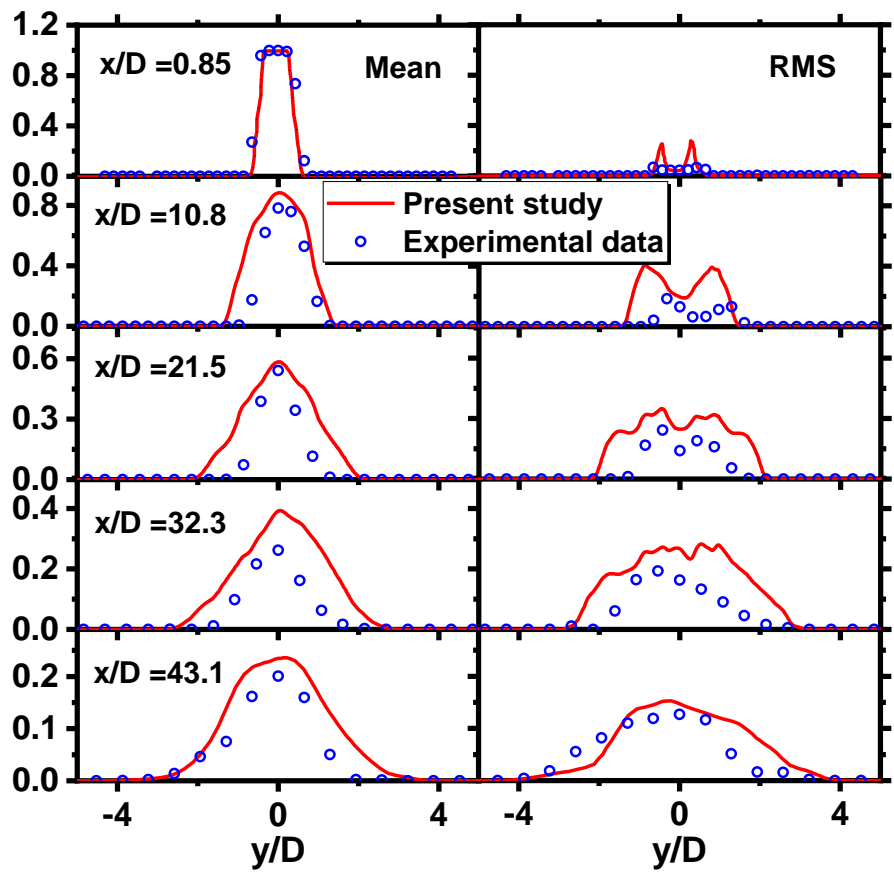

Fig. 13 Radial profiles of mean (left column) and RMS (right column) of $H_2$ mole fraction. Experimental data from Ref. [27].

The radial profiles including mean and RMS values of $H_2$ and $H_2O$ mole fractions are shown in Figs. 13 and 14 for the same five axial locations as above. In addition to some overshoots of the $H_2$ RMS values in Fig. 13, $H_2$ and $H_2O$ concentrations and fluctuations are predicted reasonably well, which is similar to those reported in Refs. [3,4]. These overshoots may be due to the over-prediction of mixture fraction, related to the uncertainties of the turbulent inlet conditions of this supersonic flame. Overall, the results in Figs. 11−14 demonstrate that the LES−PSR model can well predict the temperature and major species mole fractions in this flame.

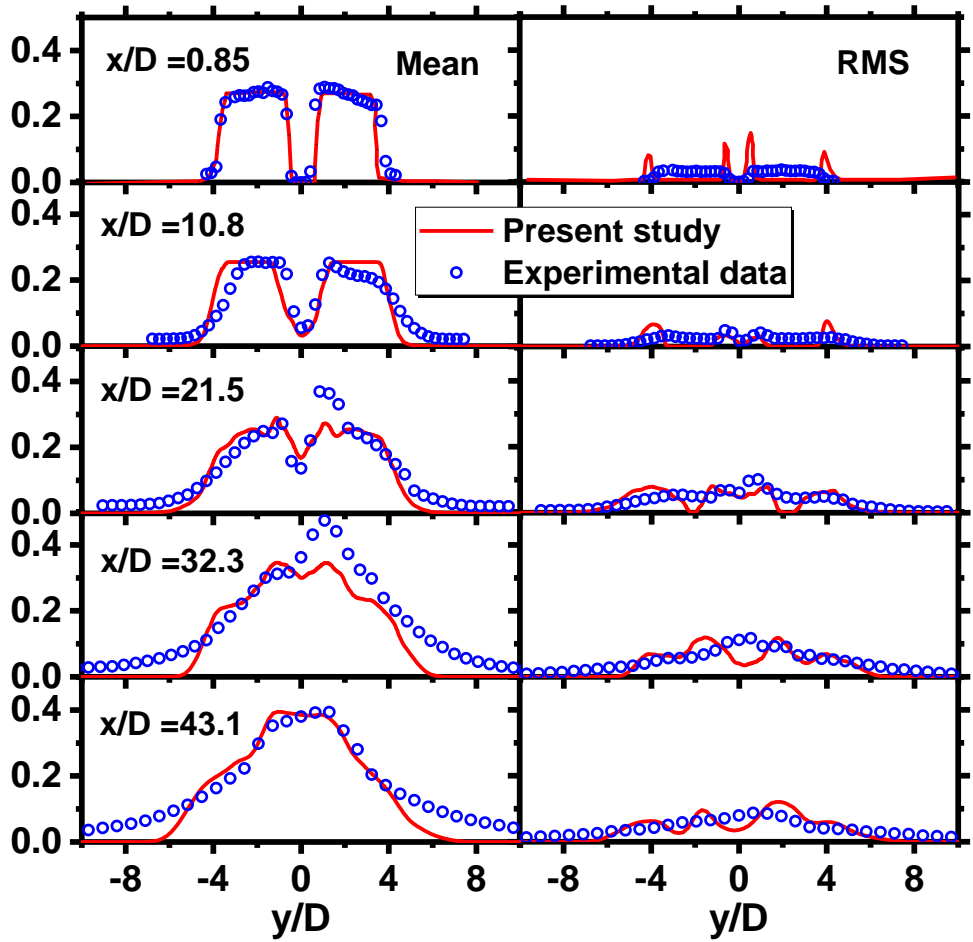

Fig. 14 Radial profiles of mean (left column) and RMS (right column) of $H_2O$ mole fraction. Experimental data from Ref. [27].

Moreover, the results from the PSR model are also compared to the those with the QLC model (see Appendix B). It is shown that the present model is more accurate in predicting reactive scalar statistics since the sub-grid scale effects are considered. Meanwhile, the computational cost is around half of that with QLC model in modelling the Cheng flame with the same numerical implementations (e.g. mesh resolution).

### 5.3 *Scalar statistics in mixture fraction space*

Scatter plots of temperature and mole fractions of $H_2$, $H_2O$ and OH against mixture fraction are presented in Figs. 15 and 16 for two locations, $(x/D, y/D) = (10.8, 0.65)$ and $(32.3, 1.1)$, respectively. The two locations are marked in Fig. 7(d), as $p_A$ and $p_B$. The first location is in the induction zone, and thus the computed temperature, $H_2$ and $H_2O$ mole fractions follow the mixing line roughly, as seen from Fig. 15. There is no salient OH at this location from the LES. However, low OH mole fraction (< 0.005) for very low mixture fractions (< 0.04) are observed in the experiments, which may be due to the fuel-lean combustion products in the hot co-flowing [27]. Moule et al. [4] observed finite OH for this location in their simulation but for a shifted (0.01−0.06) mixture fraction range. This may be because the part of the nozzle is included in Ref. [4], suggesting that the variations of OH and temperature in the induction zone are sensitive to upstream turbulence developments. Moreover, the conditional means are shown to be close to those from the experiments.

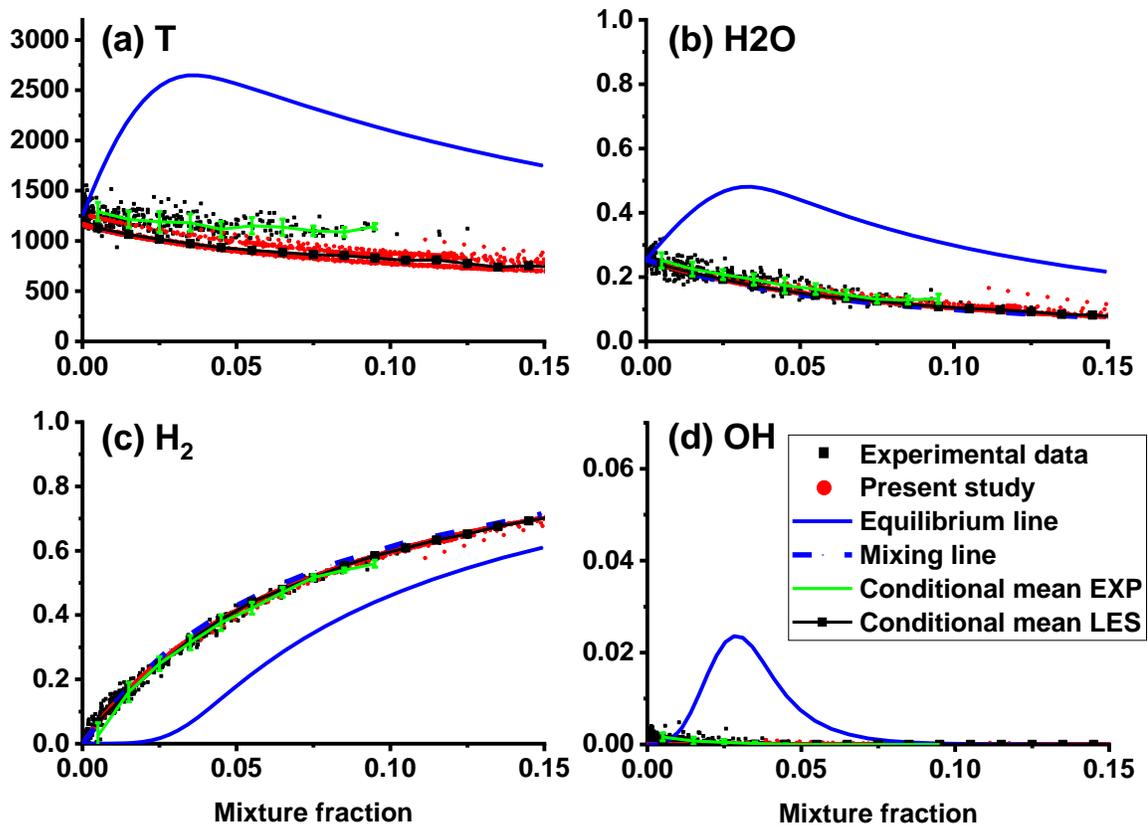

Fig. 15 Scatter plots of temperature and species mole fractions at $x/D = 10.8$ and $y/D = 0.65$. Solid blue line: equilibrium line; dashed blue line: mixing line; solid green line: conditional mean of experimental data; solid black line with symbol: conditional mean of present LES data. Red dot: LES; black dot: experimental data [27].

Figure 16 shows the scatter plot for $(x/D, y/D) = (32.3, 1.1)$, which is beyond the computed lift-off height of about $26D$ (see Fig. 7d). Overall, the simulation well captures the thermo-chemical states through comparing against the measurements. Two features are worth noting. First, the computed mixture fraction varies approximately from 0 to 0.15, larger than the measured range of 0−0.08, but it is close to the results in Refs. [4]. Second, the fluctuations in temperature and three species mole fractions are under-predicted. Also, the conditional means of temperature and $H_2O$ and OH mole fractions are slightly under-predicted in mixture fraction space, while the conditional mean of species $H_2$ is over-predicted. These under-predictions and over-predictions are also observed in Refs. [3] and [4]. They may be because the possible chemical non-equilibrium in the combustion product of $H_2$ and $O_2$-enriched air in the co-flowing jet, which is not considered in the present and previous simulations [3,4]. However, Figs. 15 and 16 have shown that the flame structures in different flame development stages are captured reasonably well.

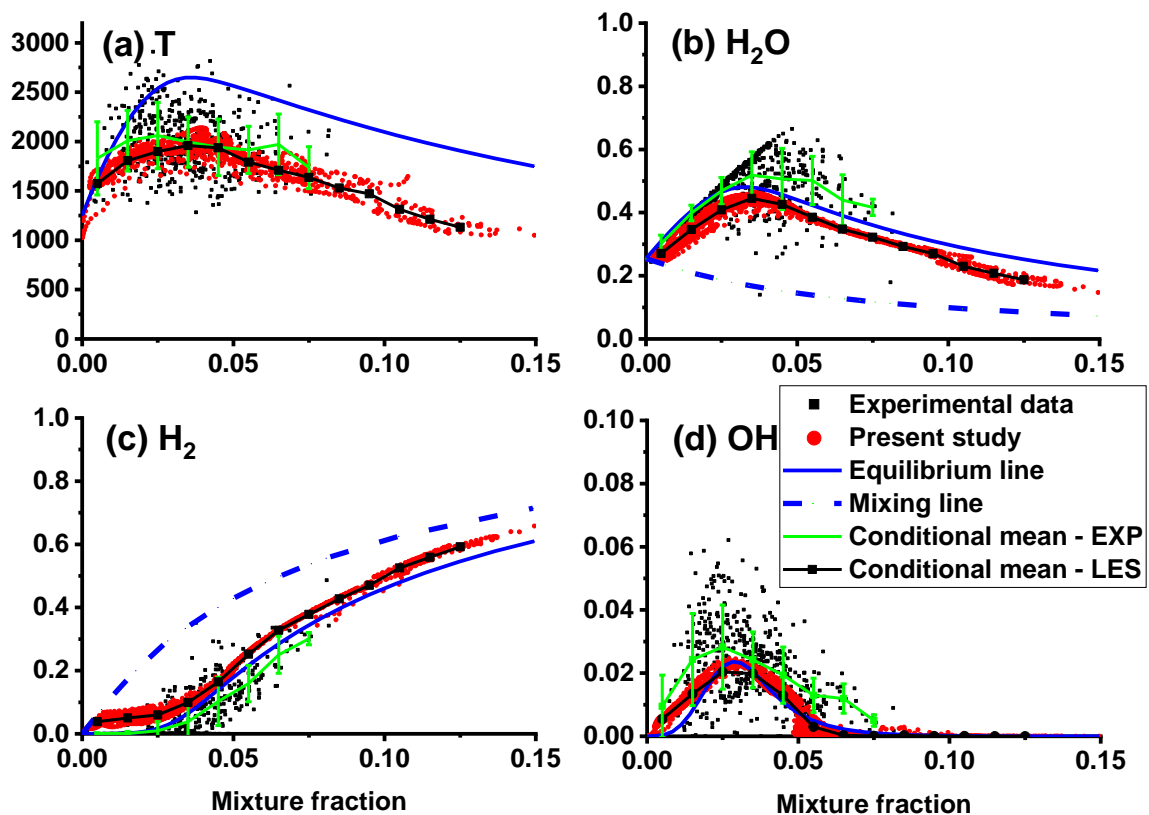

Fig. 16 Scatter plots of temperature and species mole fractions at $x/D = 32.3$ and $y/D = 1.1$. Legend same as in Fig. 15.

## 5.4 *Auto-ignition dynamics under supersonic conditions*

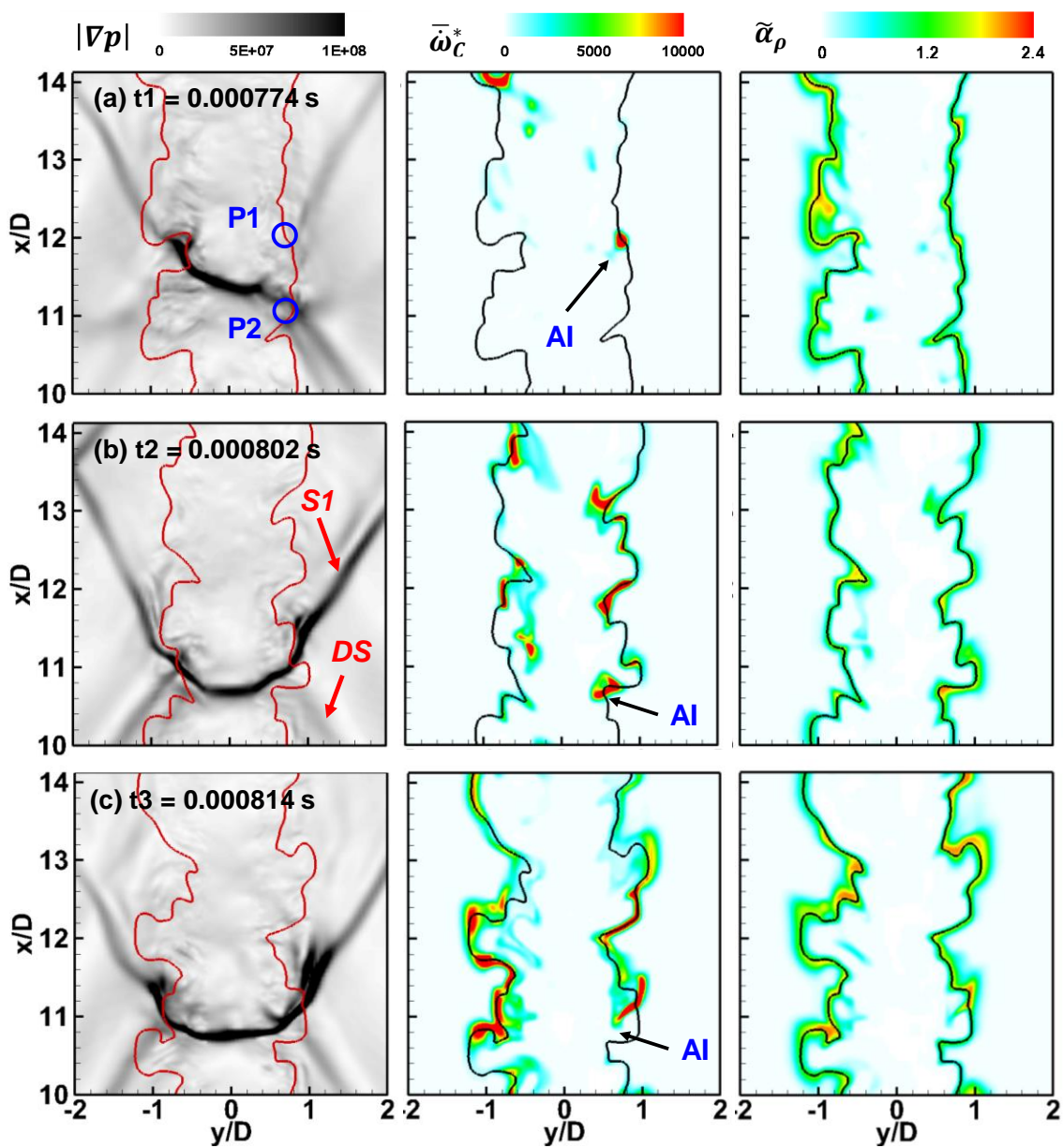

Fig. 17 Contours of pressure gradient magnitude (in Pa/m) (first column), source term (s$^{-1}$) of $\tilde{c}$ equation (second column) and density correction exponent $\tilde{\alpha}_\rho$ (third column) at three instants: (a) $t_1$ = 0.000774 s, (b) $t_2$ = 0.000802 s, (c) $t_3$ = 0.000814 s. "*DS*" represents the diamond shock and "*S1*" is V-shaped shock. P1: (12*D*, 0.75*D*, 0*D*), P2: (12*D*, 0.75*D*, 0*D*). AI: auto-ignition. Iso-lines: stoichiometric mixture fraction.

The spatial distributions of pressure gradient magnitude $|\nabla p|$, $\tilde{c}$ equation source term $\bar{\dot{\omega}}_C^*$ and density correction exponent $\tilde{\alpha}_\rho$ in the region of $10D \leq x \leq 14D$ and $-2D \leq y \leq 2D$ are shown in Fig. 17. Recall that the shock diamond ends at $10D-12D$ and a strong V-shaped shock starts there (see Fig. 7a), which is responsible for the pronounced increase of the static temperature and pressure leading to the formation of localized ignited spots. The ignited spots at birth demonstrate an intermittent topology, which are disconnected from the bulk flame zones downstream. This topology implies that the ignited spots are not induced by the upstream propagation of the bulk flame, but rather by Auto-Ignition (AI) in the upstream shocked mixtures [58]. The AI spots indicated in Fig. 17 are initiated typically in lean mixtures close to the stoichiometric mixture fraction iso-lines ($Z_{st} = 0.03$). The AI locations at the three instants indicate that the AI spots develop with the evolution of V-shaped shock and these isolated AI spots grow and are transported further downstream. This causes the flame anchoring point to fluctuate between $10D$ and $20D$. Localized autoigniting spots are also observed by Markides and Mastorakos [59] in low-speed (hence shockless) hydrogen jet in vitiated flows. Different from our results, their AI kernels are more random, which are attributed to the turbulence and air temperature fluctuations. Moreover, it should be noted that the density correction exponent $\tilde{\alpha}_\rho$ around the AI spot along the right branch of the iso-lines is high (over 2.0). This also indicates that the enhanced chemical reactions immediately behind the shock are captured in the PSR model.

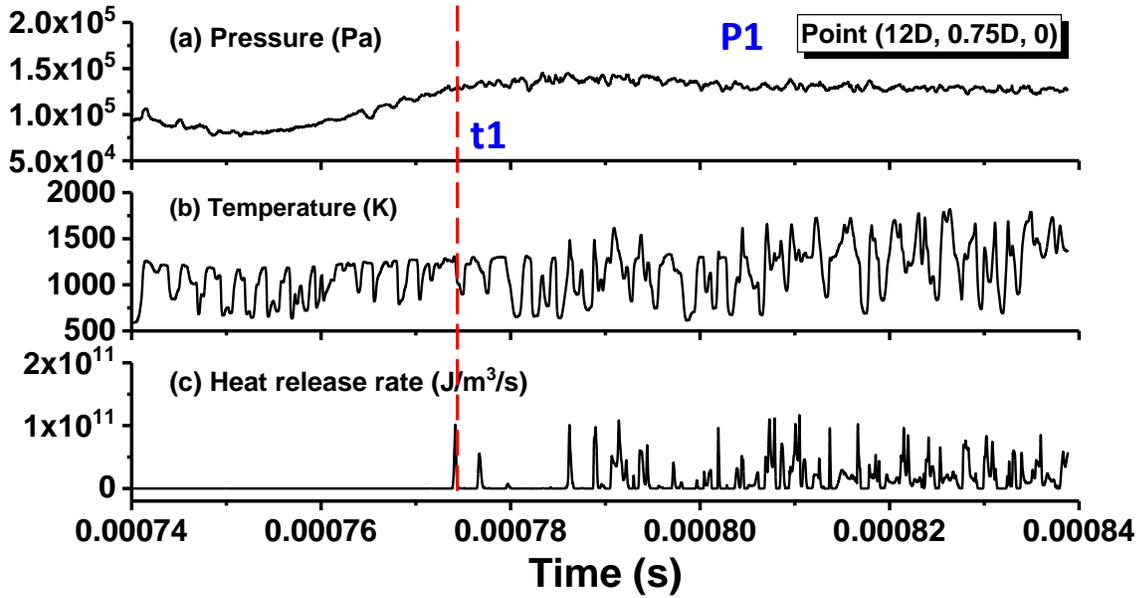

Fig. 18 Time history of (a) pressure (in Pa), (b) temperature (K) and (c) and heat release rate (J/m³/s) at probe P1 shown in Fig. 17(a).

To elaborate the evolutions of the AI spots, Figs. 18 and 19 show the time histories of pressure, temperature and heat release rate at two probes shown in Fig. 17(a), i.e., P1 (12$D$, 0.75$D$, 0$D$) and P2 (12$D$, 0.75$D$, 0$D$). They correspond to two representative scenarios of AI spot formation. Specifically, at P1, an AI spot (see Fig. 17a) formed at $t_1$ is due to the evolution of the V-shaped shock wave (see Fig. 17a). However, at P2, the AI spots at $t_2$ and $t_3$ (see Figs. 17b and 17c) are formed because of sudden pressure elevation due to shock wave movement. It can be found that considerable pressure changes are observed at the two probe locations, ranging from 0.8 to 2.0 atm, which significantly affect the AI development. As seen in Figs. 18 and 19, no heat release rate is observed before 0.00077 s with the pressure less than 1.1 atm, and the peak heat release rate is observed with increased pressure. Three instants in Fig. 17 are respectively marked in Figs. 18 or 19. At $t_1$, the pressure from P1 has increased to about 1.4 atm due to the evolution of the V-shaped shock wave *S1* towards upstream of P1 and the temperature is accordingly increased to about 1,300K. Therefore, an AI spot is formed at P1 with strong

HRR, as seen in Fig. 17(a). Then the AI spot is transported downstream, manifested by rapidly reduced HRR at this location. At instant $t_2$ and $t_3$, the locally increased pressure at P2 is observed due to the interactions between the V-shaped shock wave and the diamond shock. The evolution of the unstable jet shear layer would also interact with these shock waves and lead to the movement of the shock waves, therefore leading to considerable pressure variation at this location.

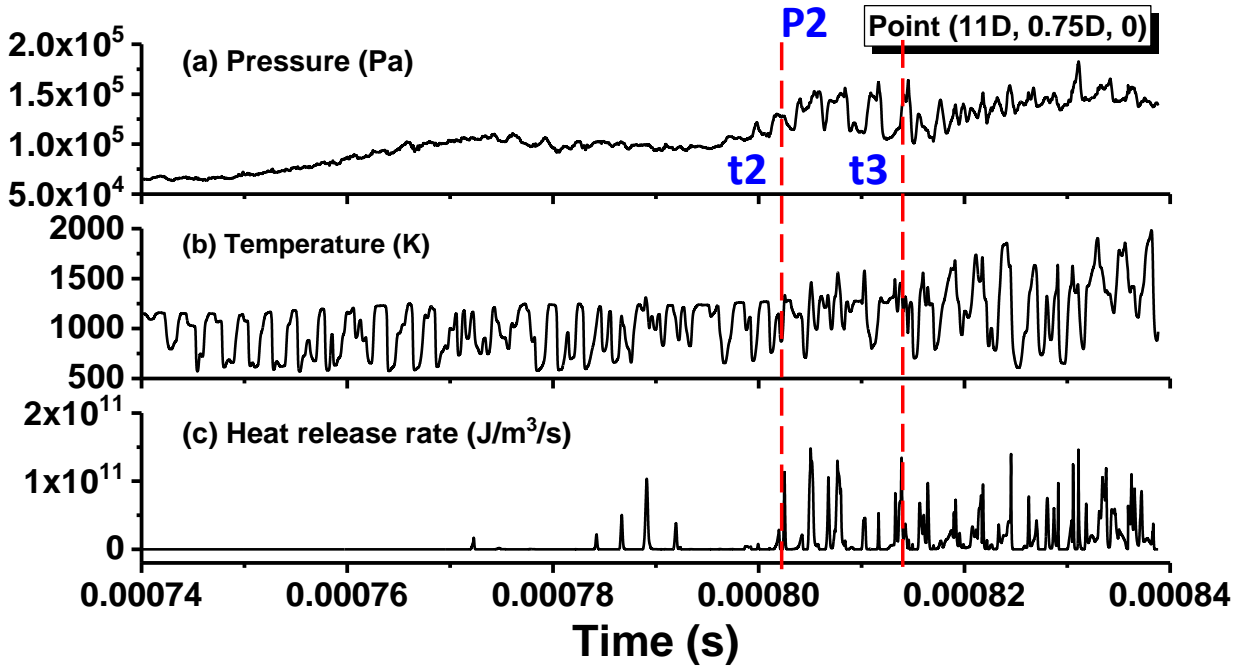

Fig. 19 Time history of (a) pressure (in Pa), (b) temperature (K) and (c) and heat release rate ($J/m^3/s$) at probe P2 shown in Fig. 15(a).

Figure 20 shows the scatters of pressure versus mixture fraction (and also equivalence ratio) which are extracted from a domain of $10 < x/D < 12$ and $(y/D)^2 + (z/D)^2 < 4$ and a duration of 0.2 ms. The scatters are colored by the source term of the $\tilde{c}$ equation. The reactive range of mixture fraction, parameterized by finite values of $\bar{\dot{\omega}}_C^*$, shown in Fig. 19 is about from 0.002 to 0.15, which is quite close to the flammable range of 0.003−0.164 suggested by Glassman [60]. It is also found that the most intense reactions (with high $\bar{\dot{\omega}}_C^*$) occur with pressure greater than 1.5 atm, where most of the mixture fractions are around the stoichiometry or under fuel-lean values. This further shows that the initiation of the chemical reactions is associated with local elevated pressure due to the shock compression.

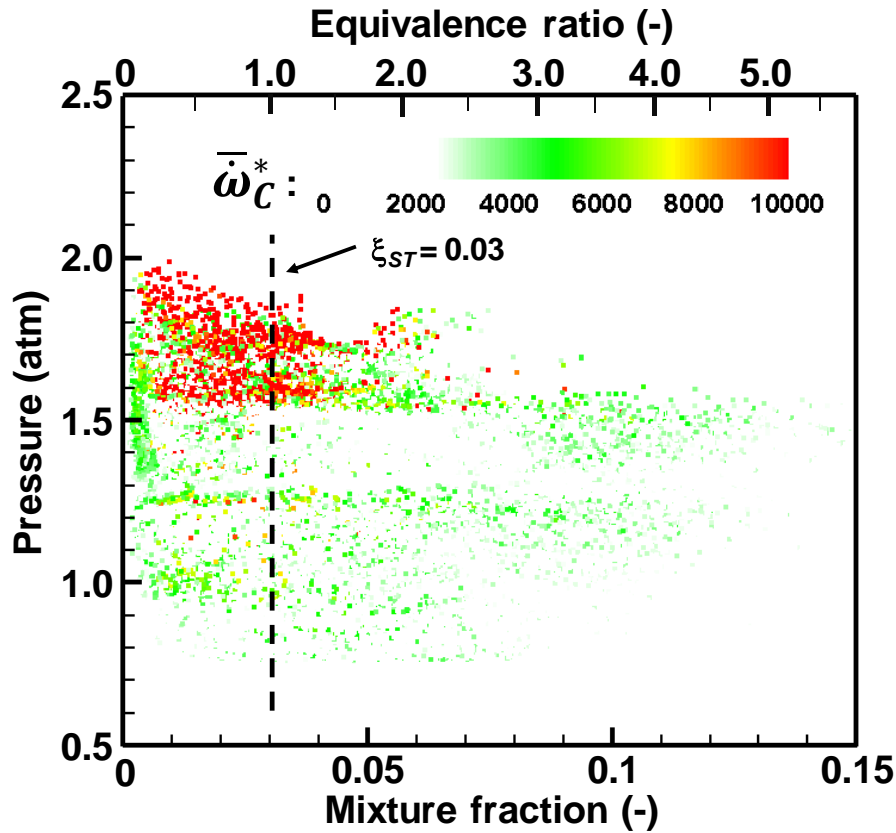

Fig. 20 Scatter plot of pressure versus mixture fraction, colored by the source term of $\tilde{c}$ equation (in kg/m$^3$/s). Data from the cylindrical domain of $10 < x/D < 12$ and $(y/D)^2 + (z/D)^2 < 4$. Dashed line: stoichiometry.

Figure 21 shows the ignition delay time versus mixture fraction (and also equivalence ratio) at different pressures. These results are calculated with a stand-alone 0D-CMC solver [61] assuming that the scalar dissipation rate is zero, as suggested by Mastorakos [62]. It is equivalent to a series of parallel PSR (homogeneous autoignition) calculations for mixture fraction range of 0−1. The thermo-chemical conditions at $Z = 0$ and $Z = 1$ are specified following the conditions of co-flowing and fuel jets listed in Table 1, respectively. The initial conditions between $0 < Z < 1$ are given as inert mixing solutions following the foregoing boundaries. One can see from Fig. 21 that the ignition delay time varies non-monotonically with mixture fraction, consistent with what has been presented by Kerkemeier et al. [63]. Specifically, when the pressure is low, e.g., 0.5 atm, the ignition delay time decreases first and then increases with the mixture fraction and has the smallest value at the stoichiometry. Therefore, the most reactive mixture fraction $\xi_{MR}$ (marked as A in Fig. 21) is equal to the stoichiometry $\xi_{ST}$ for the pressure of 0.5 atm. However, different from the results by Kerkemeier et al. [63], the shortest ignition delay time can be achieved through distributed mixture fractions (marked as B−D) when the pressure is further increased, and accordingly the most reactive mixture fraction range is significantly extended with pressure. For instance, as seen in Fig. 21, the most reactive mixture fraction is 0.023−0.03 at 1 atm, whilst 0.017−0.03 at 2 atm and 0.013−0.03 at 3 atm. Apparently, this extension always occurs under the fuel-lean conditions for hydrogen and air mixtures considered here. Meanwhile, the shortest ignition delay time decreases considerably with pressure, i.e., about 0.047, 0.018, 0.008 and 0.005 ms, corresponding to 0.5−3 atm in Fig. 21. Therefore, the elevated pressure effects on the most mixture fraction and shortest ignition delay time suggest that combustion would occur favorably under distributed fuel-lean mixture compositions with increased pressure. This is a novel feature in supersonic combustion, which has not been reported by Mastorakos [62] (simply mentioning that "*the effect of pressure on $\xi_{MR}$ has not been studied yet*"). This well justifies the distributed combustion in supersonic H$_2$ flame at elevated pressures observed in Fig. 20. Similar pressure influences on $\tau_{ig}$ and $\xi_{MR}$ are also present in our recent LES of supersonic ethylene flames [48], and the distributed reaction layer is experimentally observed by Gamba and Mungal [22] from OH-PLIF images of hydrogen combustion in supersonic crossflows.

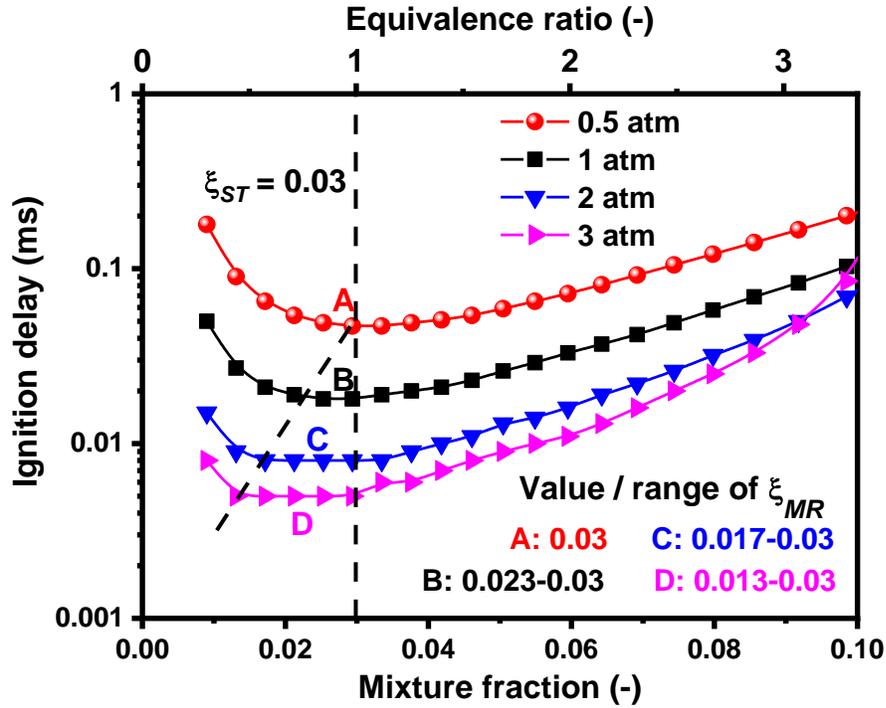

Fig. 21 The ignition delay time versus mixture fraction at different pressures. $\xi_{ST}$: stoichiometric mixture fraction; $\xi_{MR}$: most reactive mixture fraction.

## 6. Conclusions

The LES of a sonic auto-ignition-stabilized hydrogen flame with supersonic viatited co-flowing [27] is conducted using the PSR method to model sub-grid scale combustion and its interactions with supersonic turbulence. The flow compressibility and viscous heating effects on the thermo-chemical state in supersonic combustion are included in the PSR model, through correcting the chemical source term for the progress variable equation and incorporating absolute enthalpy as one of the control variables for the look-up table.

Firstly, *a priori* analysis of the compressibility and viscous heating effects is performed to validate the proposed PSR model for modelling supersonic combusion. The results show that both the major species, e.g. $H_2$ and $O_2$, and the minor species, e.g. OH, H and $HO_2$, are slightly affected by the initial temperature and pressure. However, the source term of the progress variable is significantly affected by the temperature and pressure. Our results from *a priori* analysis have shown that the developed PSR model can accurately describe the response of the thermo-chemical state to the elevated temperature and pressure which arise from shock compression in supersonic combustion.

The LES−PSR method is then validated in a sonic auto-ignition-stabilized turbulent hydrogen flame with supersonic viatited co-flowing [27]. The results show that the near-field shock diamonds, overall flame characteristics, flame-shock interaction and lift-off height are well predicted. The velocity statistics show fairly good agreement with the measurements. The mean and RMS mixture fractions are captured quite well except for some over-predictions at an upstream location, which may be related to approximations of the inflow boundary conditions. This study also shows that the LES−PSR method can capture the mean temperature and major species mole fractions in flame induction and stabilization zones. However, under-predictions of temperature RMS is observed, which may be due to the chemical non-equilibrium in the combustion product of $H_2$ and $O_2$-enriched air in the co-flowing jet. There are also good agreements between the computed and measured flame sturctures in mixture fraction space.

The shock-induced auto-igniting spots are captured by the PSR model, and the intermittent spots play important roles in flame stabilization. Moreover, in the flame stabilization region, it is shown that the mixture fractions with intense reaction are mainly around the stoichiometry or under fuel-lean values with elevated pressure due to shock compression. Calculations of most reactive mixture fraction are performed to assess the effects of pressure on the ignition delay time. The results suggest that combustion would occur favorably under a range of fuel-lean mixture compositions with increased pressure, which well justifies the distributed combustion observed in supersonic $H_2$ flame when the pressure increases.

Finally, the advantages of the proposed combustion model for supersonic combustion are further addressed (see Appendix B). It is shown that the LES with the present combustion model shows much better performance than that with the quasi-laminar chemistry. Since the species and the reaction source terms are obtained by looking up the flamelet table, it can greatly reduce the computational cost.

**Acknowledgement**

This work used the computational resources from National Supercomputing Center, Singapore (https://www.nscc.sg/). HZ and MZ are sponsored by Singapore Ministry of Education Tier 1 grant (R-265-000-653-114). ZXC and NS are supported by Mitsubishi Heavy Industries. Professors T.-S. Cheng and R. W. Pitz are acknowledged for sharing the experimental data and burner details. Dr Zhiwei Huang from NUS is thanked for assistance with setting up the inlet turbulent conditions.

*Appendix A Inlet turbulent effects*

To further study the inlet turbulence effects, LES with synthetic turbulence [64] is repeated. The Reynolds stress is given following Ref. [65], and the integral length scales for the fuel and co-flowing jets are 0.236 mm and 1.397 mm, respectively. Figure A1 shows the distributions of pressure gradient magnitude with white noise and synthetic turbulence at the inlet. It can be found that the diamond shock structures are blurred when the turbulence is included.

Figures A2 – A4 show the profiles of mean axial velocity, mean and RMS of mixture fraction and temperatures, respectively. In general, both results show good agreement with the experimental data. However, there are still some differences. For instance, in Fig. A2 the mean axial fuel jet velocity is slightly overestimated at $x/D$ = 21.5 and 32.3. This may be due to that the strong shock waves are weakened by inlet turbulence (see Fig. A1), and therefore the effect of pressure gradient on the mean axial velocity would decrease behind the shock along the centerline. Moreover, the mixing between the fuel jet and the co-flowing is enhanced with the inlet turbulence, because the mixture fraction is significantly reduced in the near field ($x/D$ = 10.8 and 21.5), as demonstrated in Fig. A3. Due to the enhanced mixing between the fuel jet and co-flowing, the mean temperature in the jet region at $x/D$ = 10.8 is also over-predicted, while in the downstream at $x/D$ = 21.5, the mean temperature is under-predicted compared to the experimental data and the LES results without synthetic turbulence. Also, the RMS values of temperature in the near field, e.g., $x/D$ = 10.8 and 21.5, are under-predicted in the fuel jet region.

However, the enhanced mixing in the near field ($x/D$ < 21.5) does not decrease the flame lift-off height (not shown here); on the contrary, the lift-off height (29$D$) is larger than the one (about 26$D$) with white noise inlet. Both are slightly higher than the measured value in the experiment (about 25$D$).

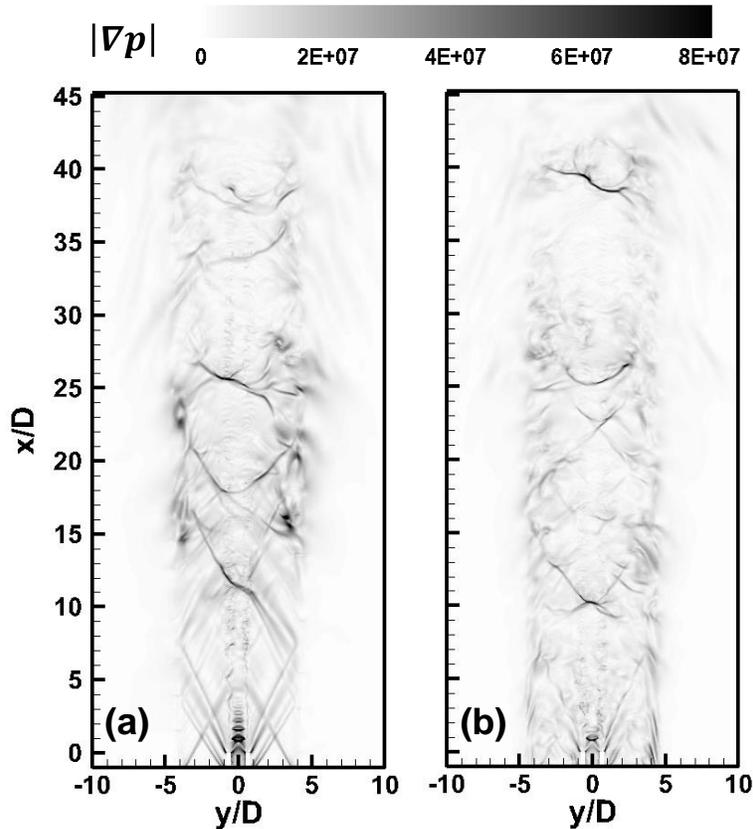

Fig. A1 Distributions of pressure gradient magnitude (in Pa/m) with (a) white noise and (b) synthetic turbulence at the inlet.

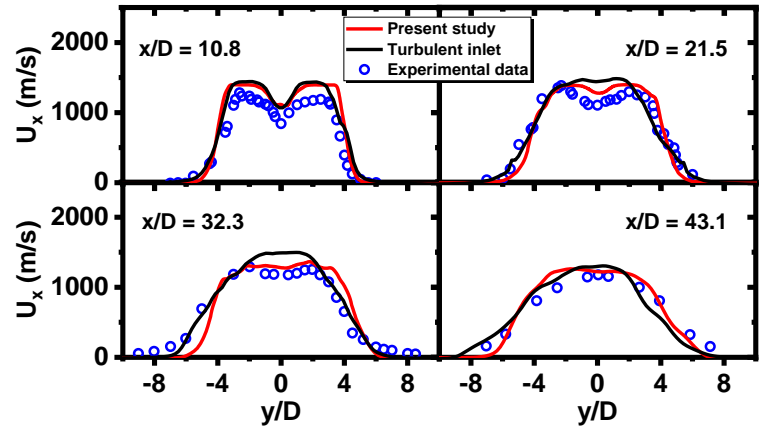

Fig. A2 Radial profiles of mean axial velocity. Comparison with experimental data from Ref. [27] and LES data with synthetic turbulence [64].

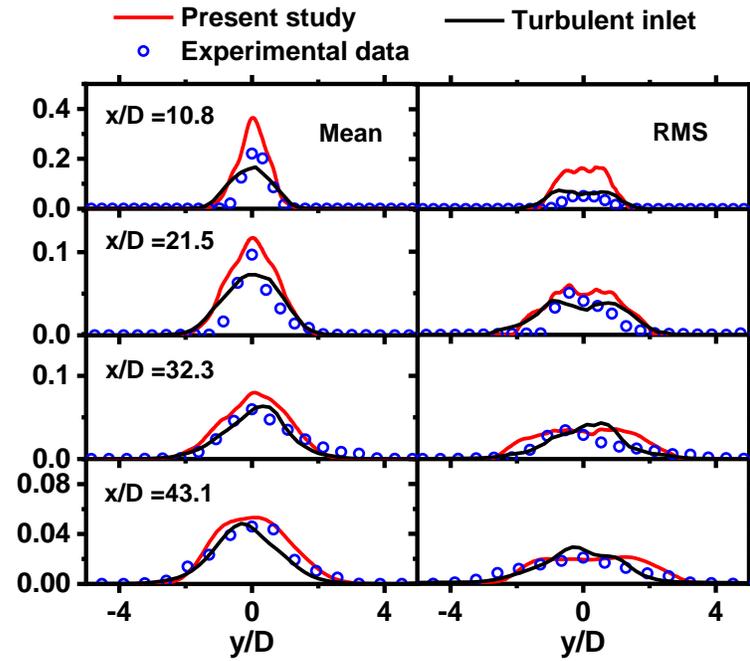

Fig. A3 Radial profiles of mean (left column) and RMS (right column) of mixture fraction. Experimental data from Ref. [27] and LES data with synthetic turbulence [64].

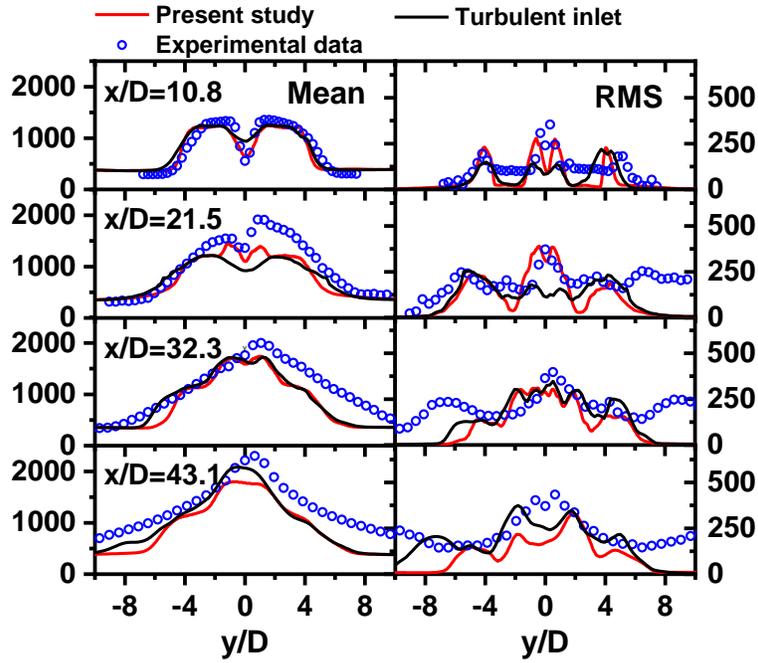

Fig. A4 Radial profiles of mean (left column) and RMS (right column) of temperatures. Experimental data from Ref. [27] and LES data with synthetic turbulence [64].

## Appendix B Comparison with the quasi-laminar chemistry model

Figures A5 – A7 compare the results with the PSR and QLC models [49]. Note that all the numerical settings, except the combustion model, are identical in both studies. In general, the flow field and flame structures are reproduced with both models. However, the temperature (Fig. A5) and $H_2O$ mole fraction (Fig. A6) along the centerline especially in the downstream are underestimated with the QLC model, where the sub-grid contribution towards the reaction rate estimation is neglected. Also, the $H_2$ mole fraction (Fig. A7) is significantly overestimated by about 5 - 57% with the QLC model. However, the LES-PSR model give much better results in the foregoing predictions.

In terms of the computational efficiency, since the species and the reaction source terms are obtained by looking up the flamelet table in LES-PSR model, the present calculations are twice as fast as the QLC modelling, with the same mechanism (9 species and 19 reactions), numerical settings and computing platform. Moreover, when more complex hydrocarbon fuels (hence more species) are involved, LES-PSR modelling is expected to be more compelling, because of more appreciably cost reduction.

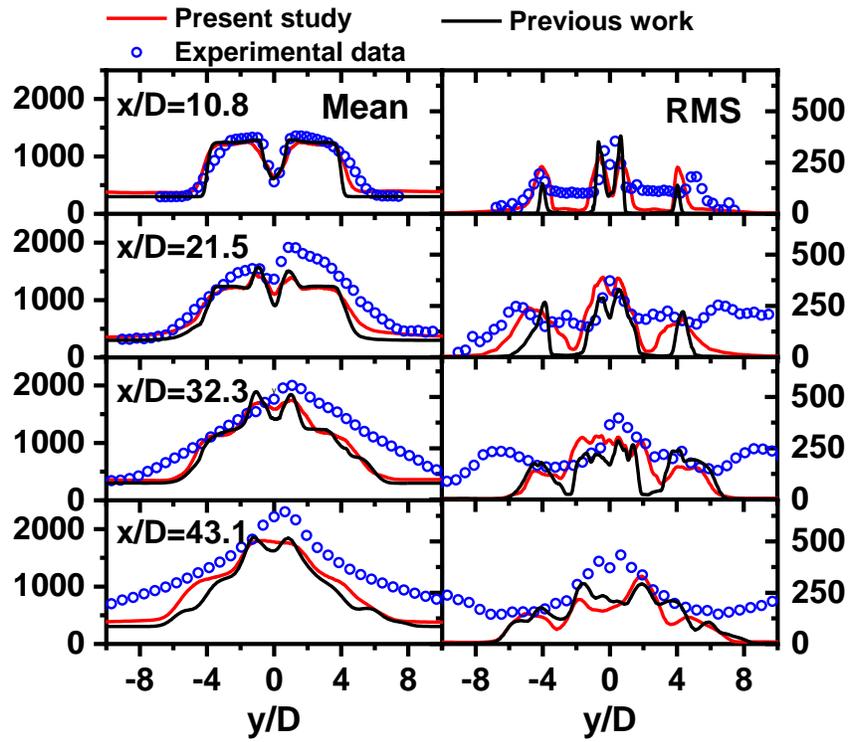

Fig. A5 Radial profiles of mean (left column) and RMS (right column) of temperatures (in K). Comparison with experimental data from Ref. [27] and LES data from Ref. [49].

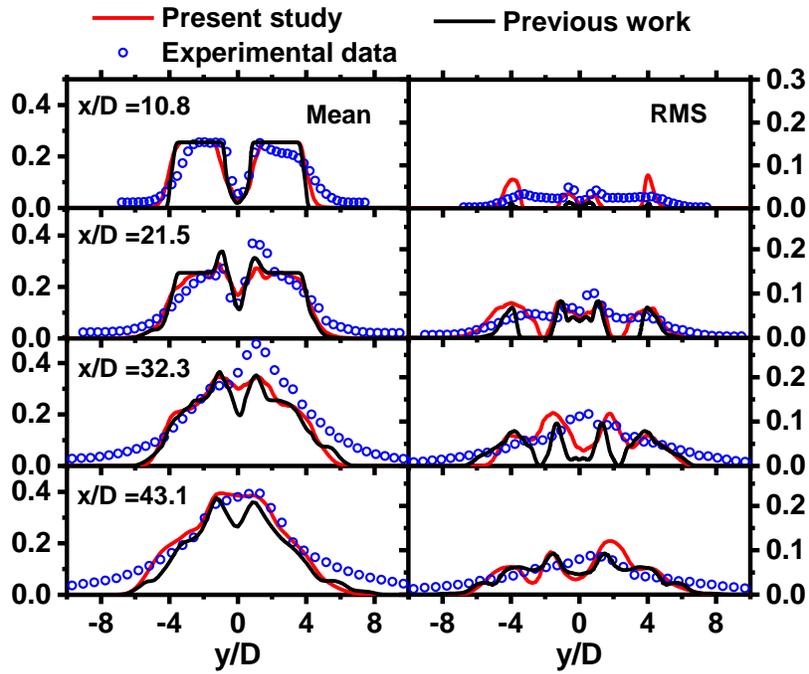

Fig. A6 Radial profiles of mean (left column) and RMS (right column) of $H_2O$ mole fraction. Comparison with experimental data from Ref. [27] and LES data from Ref. [49].

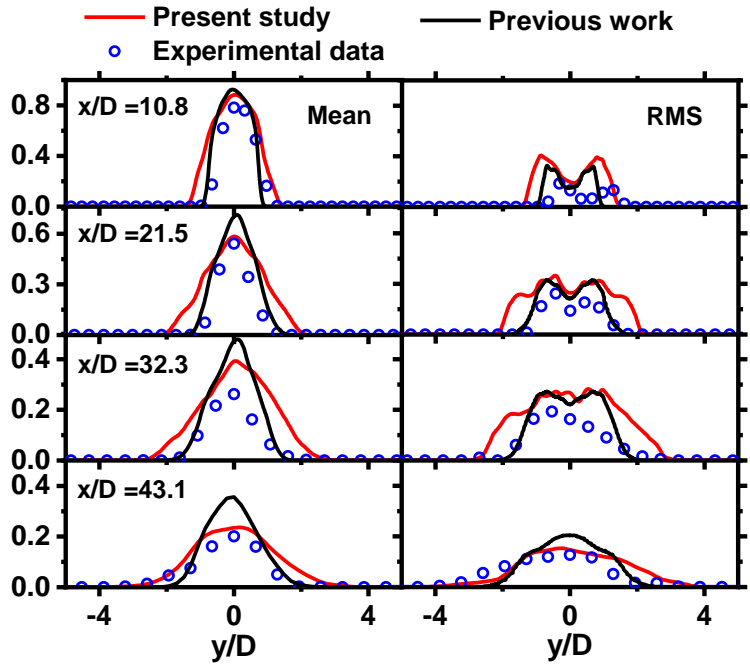

Fig. A7 Radial profiles of mean (left column) and RMS (right column) of $H_2$ mole fraction. Comparison with experimental data from Ref. [27] and LES data from Ref. [49].